\documentclass[twocolumn]{aastex63}
\usepackage{empheq}

\providecommand{\etal}{{\it et al.} }			
\providecommand{\kms}{km s$^{-1}$}			

\shorttitle{H$\alpha$ Dots II}
\shortauthors{Salzer et al.}

\begin{document}


\title{The H$\alpha$ Dots Survey.  II. A Second List of Faint Emission-Line Objects}


\correspondingauthor{John J. Salzer}
\email{josalzer@indiana.edu}

\author[0000-0001-8483-603X]{John J. Salzer}
\affiliation{Department of Astronomy, Indiana University, 727 East Third Street, Bloomington, IN 47405, USA}

\author[0000-0003-3810-3323]{Jesse R. Feddersen}
\affiliation{Department of Astronomy, Indiana University, 727 East Third Street, Bloomington, IN 47405, USA}
\affiliation{Department of Astronomy, Yale University, P.O. Box 208101, New Haven, CT 06520, USA}

\author{Kathryn Derloshon}
\affiliation{Department of Astronomy, Indiana University, 727 East Third Street, Bloomington, IN 47405, USA}

\author{Caryl Gronwall}
\affiliation{Department of Astronomy \& Astrophysics, Pennsylvania State University, University Park, PA 16802, USA}
\affiliation{Institute for Gravitation \& the Cosmos, Pennsylvania State University, University Park, PA 16802, USA}

\author{Angela Van Sistine}
\affiliation{Department of Astronomy, Indiana University, 727 East Third Street, Bloomington, IN 47405, USA}
\affiliation{Center for Gravitation, Cosmology, and Astrophysics, University of Wisconsin-Milwaukee, 3135 N Maryland Ave 
Milwaukee, Wisconsin 53211, USA}

\author{Arthur Sugden}
\affiliation{Department of Astronomy, Indiana University, 727 East Third Street, Bloomington, IN 47405, USA}

\author[0000-0001-9165-8905]{Steven Janowiecki}
\affiliation{University of Texas at Austin, McDonald Observatory, TX 79734, USA}

\author[0000-0002-2954-8622]{Alec S. Hirschauer}
\affiliation{Space Telescope Science Institute, 3700 San Martin Dr., Baltimore, MD 21218, USA}

\author{Jessica A. Kellar}
\affiliation{Department of Physics \& Astronomy, Dartmouth College, Hanover, NH  03755, USA}

\begin{abstract}
We present the second catalog of serendipitously discovered compact extragalactic emission-line sources  -- H$\alpha$ Dots.    These objects have been discovered in searches of moderately deep narrow-band images acquired for the ALFALFA H$\alpha$ project \citep{vansistine2016}.  In addition to cataloging 119 new H$\alpha$ Dots, we also present follow-up spectral data for the full sample.   These spectra allow us to confirm the nature of these objects as true extragalactic emission-line objects, to classify them in terms of activity type (star forming or AGN), and to identify the emission line via which they were discovered.  We tabulate photometric and spectroscopic data for the all objects, and present an overview of the properties of the full H$\alpha$ Dot sample.  The H$\alpha$ Dots represent a broad range of star-forming and active galaxies detected via several different emission lines over a wide range of redshifts.  The sample includes H$\alpha$-detected blue compact dwarf galaxies at low redshift, [\ion{O}{3}]-detected Seyfert 2 and Green Pea-like galaxies at intermediate redshifts, and QSOs detected via one of several UV emission lines, including Ly$\alpha$.  Despite the heterogeneous appearance of the resulting catalog of objects, we show that our selection method leads to well-defined samples of specific classes of emission-line objects with properties that allow for statistical studies of each class.
\end{abstract}




\section{Introduction} 

Much of what we currently know about activity in galaxies -- either extreme star-formation events or non-stellar emission from Active Galactic Nuclei (AGN) -- has been learned by studying systems originally recognized in focused surveys for such activity.   These surveys have taken place at essentially all wavelengths, from the radio to X-rays.  Ground-breaking surveys include the early 3C radio continuum surveys \citep{edge1959, bennett1962, laing1983} and the Markarian objective-prism searches for UV-excess galaxies \citep[e.g.,][]{mrk67, mrk81}.

Of particular relevance to the current study are the many surveys that have cataloged galaxies with strong optical emission lines, so called emission-line galaxies (ELGs).  ELG surveys have tended to use one of two selection methods, although there are many variations on these two primary themes.  The first utilizes objective-prism searches using wide-field Schmidt telescopes \citep[e.g.,][]{UM1977, UM1981, Case82, Case83, SBS1, Was83, UCM94, UCM96, HS1999, HS2000, salzer2000, salzer2001}.  A key advantage of the objective-prism method is that it yields wide-field data, covering substantial area in reasonable amounts of time.  In addition, the spectral coverage of the prism typically allows for the detection of ELGs over a wide redshift range in a single exposure.  

The second method involves the use of narrow-band filters to isolate the strong line emission in ELGs.  Examples of surveys that utilize this latter method include \citet{BST1993, Kakazu2007, cook2019}.  Narrow-band surveys typically employ smaller fields-of-view than objective-prism surveys, and are limited in their spectral range to the bandwidth of the filter or filters employed.  They do, however, often result in deeper, more sensitive searches for ELGs.

The ELG survey presented in this study is of the latter type: narrow-band detection.  We refer to our program as the H$\alpha$ Dots survey \citep{hadot1}, since we are searching for compact/unresolved sources of emission in images taken through narrow-band H$\alpha$ filters.  The H$\alpha$ Dots survey is carried out utilizing a large set of narrow-band data obtained as part of a completely different program: the ALFALFA H$\alpha$ project \citep[e.g.,][hereafter AHA]{vansistine2016}.  The H$\alpha$ Dots program originated with the serendipitous discovery of unresolved emission-line sources in the AHA data.  This initial effort resulted in our first survey list of 61 compact emission-line sources \citep{hadot1}.

The current paper presents the second list of H$\alpha$ Dots, discovered by searching narrow-band images from several additional AHA observing runs.  As with the first H$\alpha$ Dots catalog, most of the work to analyze and select the faint emission-line sources was carried out by undergraduate students working with the senior author.  Our paper is organized as follows: Section 2 describes the data used for the searches as well as our search methodology. Section 3 presents our new list of objects, while Section 4 details our spectroscopic follow-up of the newly-discovered ELGs.  Section 5 presents key properties of the H$\alpha$ Dots, including their redshift distribution, spectroscopic properties, object classifications, luminosity and star-formation rate distributions, and metal abundances.  Section 6 summarizes our main results.  A standard $\Lambda$CDM cosmology with $\Omega_m = 0.27$, $\Omega_\Lambda = 0.73$, and $H_0 = 70$~kms$^{-1}$ Mpc$^{-1}$  is assumed in this paper.

\section{Observational Data and Search Technique} 

As mentioned above, the H$\alpha$ Dots survey is carried out using narrow-band images obtained for the AHA project \citep{vansistine2016}.  The AHA project was a narrow-band imaging survey that observed a volume-limited sample of HI-detected galaxies cataloged in the ALFALFA survey \citep{giovanelli2005, haynes2011, haynes2018}.   The primary goal of the AHA project was to study the star-formation properties of a comprehensive sample of galaxies in the local universe, using the deep ALFALFA survey catalog as a source list.   A total of 1555 galaxies were observed.

The AHA data consist of images obtained through both a continuum R-band filter and one of a set of narrow-band ($\sim$60-70 \AA\ wide) H$\alpha$ filters.  Two different narrow-band filter sets were employed \citep[see][]{vansistine2016}, appropriate for galaxies with velocities up to $\sim$9200 \kms.  Since the ALFALFA targets all have accurate redshifts from the HI detections, the appropriate narrow-band filter to use for each object was always clear.   

Two telescopes were used for the AHA observations: the WIYN 0.9-m\footnote{The 0.9 m telescope is operated by WIYN Inc. on behalf of a Consortium of nine partner Universities and Organizations (see www.noao.edu/0.9m/ partners).  WIYN is a joint partnership of the University of Wisconsin at Madison, Indiana University, and the National Optical Astronomical Observatory.} and KPNO 2.1-m telescopes\footnote{The KPNO 2.1-m telescope was formerly operated by the National Optical Astronomy Observatory (NOAO), which consisted of KPNO near Tucson, Arizona, Cerro Tololo Inter-American Observatory near La Serena, Chile, and the NOAO Gemini Science Center. NOAO was operated by the Association of Universities for Research in Astronomy (AURA) under a cooperative agreement with the National Science Foundation}.  The current paper deals exclusively with images obtained with the WIYN 0.9-m telescope.  All images were acquired with the S2KB CCD detector.   We read out only the central 1536 $\times$ 1536 pixels, which yielded a field-of-view of $\sim$15.2 arcmin square.  A pair of 1200 s exposures obtained through the appropriate narrow-band filter were sandwiched around a 240 s R-band image for each target galaxy.  Standard image processing procedures were applied to the data, after which the three images for a given galaxy were run through a custom pipeline to assign a world coordinate system and to scale and subtract the continuum R-band image from the narrow-band images \citep[see][]{vansistine2016}.

Accurate flux calibration was essential for the AHA project.  Multiple observations of spectrophotometric standard stars were carried out on each night of the program.   On nights when conditions were not strictly photometric, fields that were observed were flagged and re-observed on a later night using short ``post-calibration" observations.   All narrow-band fluxes measured from AHA images have the zero-point of the flux scale measured to better than 2\% accuracy (and usually $\sim$1\%).  Full details are included in \citet{vansistine2016}.

The H$\alpha$ Dot survey uses the fully processed images from the AHA project to search for point-like objects with an excess of emission in the narrow-band filter images.  The search process is fully automated, and is described in detail by \citet{hadot1}.   In brief, the images for each field are scanned to identify all of the objects present in the field, after which the fluxes for each source are measured through the same small aperture in both the R-band and narrow-band images.  Once the candidate H$\alpha$ Dots have been identified by our software, each potential dot is examined critically in the imaging data to weed out false detections (e.g., image artifacts).

\begin{deluxetable*}{ccccrccccc}
\tabletypesize{\footnotesize}
\tablewidth{0pt}
\tablecaption{Second List of H$\alpha$ Dots \label{tab:prop}}

\tablehead{
 \colhead{H$\alpha$ Dot \#} &  \colhead{RA} &  \colhead{DEC} &  \colhead{$\Delta$m} &  \colhead{Ratio} & \colhead{m$_R$} & \colhead{NB Line Flux} & \colhead{M$_R$} &  \colhead{log(L(H$\alpha$))} &  \colhead{log(SFR)}  \\
  &  \colhead{degrees} &  \colhead{degrees} &  &  &  & x10$^{-14}$ erg/s/cm$^2$ &  &  erg/s  & \colhead{M$_\odot$/yr} \\
 (1)  & (2)  & (3)  & (4)  & (5)  & (6)  & (7)  & (8)  & (9)  & (10)
}
 \startdata
   62 &  138.33285 &   12.53536 &  -0.83 &  22.07 &  17.74 $\pm$ 0.04 & 1.030 $\pm$ 0.054 & -16.53 &  39.80 & -1.30 \\
   63\tablenotemark{a} &  138.80328 &   11.88416 &  -1.54 &   5.00 &  20.96 $\pm$ 0.27 & 0.279 $\pm$ 0.032 & -13.36 &  39.27 & -1.83 \\
   64 &  138.84261 &   11.83980 &  -0.34 &   6.37 &  19.36 $\pm$ 0.08 & 0.187 $\pm$ 0.024 &  -- & -- & -- \\
   65 &  139.62692 &   13.77032 &  -1.52 &  10.35 &  20.04 $\pm$ 0.17 & 0.325 $\pm$ 0.028 & -21.20 &  42.14 & \ 1.04 \\
   66 &  147.92081 &   14.10914 &  -0.63 &  13.92 &  18.91 $\pm$ 0.08 & 0.336 $\pm$ 0.029 & -14.21 &  38.76 & -2.34 \\
   67 &  149.25146 &   15.55564 &  -0.56 & 111.30 &  15.41 $\pm$ 0.03 & 5.982 $\pm$ 0.060 & -18.58 &  40.94 & -0.17 \\
   68 &  149.65731 &   15.43306 &  -1.82 &  17.51 &  20.12 $\pm$ 0.19 & 0.419 $\pm$ 0.032 &  -- & -- & -- \\
   69 &  149.67200 &   15.40155 &  -0.71 &  13.30 &  18.15 $\pm$ 0.08 & 0.427 $\pm$ 0.027 & -16.01 &  40.12 & -0.98 \\
   70 &  159.09391 &   13.63132 &  -0.87 &  14.70 &  20.05 $\pm$ 0.05 & 0.191 $\pm$ 0.015 & -21.19 &  42.52 & -- \\
   71 &  171.07641 &   14.31403 &  -0.40 &   6.58 &  18.85 $\pm$ 0.09 & 0.168 $\pm$ 0.030 & -22.44 &  40.83 & -- \\
\\   
   72 &  174.32659 &   15.50637 &  -1.05 &  22.92 &  19.26 $\pm$ 0.05 & 0.518 $\pm$ 0.021 & -22.05 &  42.86 & \ 1.75 \\
   73\tablenotemark{a} &  175.48027 &   15.94398 &  -0.63 &   6.10 &  19.29 $\pm$ 0.10 & 0.196 $\pm$ 0.047 & -21.89 &  42.20 & \ 1.10 \\
   74 &  177.07405 &   12.71669 &  -2.23 &   9.59 &  21.25 $\pm$ 0.25 & 0.371 $\pm$ 0.025 & -12.48 &  39.60 & -1.51 \\
   75 &  181.35257 &   15.41589 &  -0.82 &   6.11 &  20.64 $\pm$ 0.12 & 0.091 $\pm$ 0.019 & -27.55 & -- & -- \\
   76 &  209.48133 &   14.43082 &  -1.46 &  21.46 &  20.75 $\pm$ 0.06 & 0.243 $\pm$ 0.013 & -20.51 &  42.55 & \ 1.45 \\
   77 &  349.83866 &   26.13977 &  -0.96 &  79.89 &  16.93 $\pm$ 0.03 & 3.053 $\pm$ 0.061 & -17.07 &  40.25 & -0.85 \\
   78\tablenotemark{a} &   10.15819 &   27.04192 &  -1.95 &   6.24 &  21.43 $\pm$ 0.32 & 0.263 $\pm$ 0.030 & -13.12 &  39.33 & -1.77 \\
   79 &   11.15322 &   26.92934 &  -1.35 &  51.10 &  18.43 $\pm$ 0.04 & 1.524 $\pm$ 0.042 & -16.07 &  40.07 & -1.03 \\
   80 &    1.63913 &   27.24681 &  -0.52 &   7.96 &  19.82 $\pm$ 0.05 & 0.153 $\pm$ 0.018 & -28.49 & -- & -- \\
   81 &    2.03670 &   27.45397 &  -1.66 &  46.27 &  18.85 $\pm$ 0.04 & 1.653 $\pm$ 0.042 & -15.50 &  40.12 & -0.98 \\
\enddata
\tablenotetext{a}{Indicates duplicate detections: H$\alpha$ Dot 63 = H$\alpha$ Dot 108; H$\alpha$ Dot 73 = H$\alpha$ Dot 30; H$\alpha$ Dot 78 = H$\alpha$ Dot 47; H$\alpha$ Dot 86 = H$\alpha$ Dot 51; H$\alpha$ Dot 109 = H$\alpha$ Dot 148; H$\alpha$ Dot 129 = H$\alpha$ Dot 49; H$\alpha$ Dot 160 = H$\alpha$ Dot 43} 
\tablecomments{Table 1 is published in its entirety in the machine-readable format.  A portion is shown here for guidance regarding its form and content.}

\end{deluxetable*}

Three selection criteria are used to identify potential H$\alpha$ dots.  First, the narrow-band image must show an excess of flux over that measured in the continuum (R-band) image such that
\begin{equation}
    \Delta m = m_{NB} - m_R \leq -0.4.
\end{equation}
Here m$_{NB}$ is the instrumental magnitude measured in the narrow-band filter, which includes both the line and continuum emission.  This value of $\Delta$m corresponds to f$_{H\alpha}$/f$_{cont}$ $\approx$ 0.45.   Given that the typical narrow-band filter used for the AHA project has a FWHM of $\sim$70 \AA, this implies an emission-line equivalent width limit for detection of $\sim$30 \AA.  Second, the ratio of the narrow-band flux excess to the formal uncertainty in that excess (both measured in magnitudes) must be
\begin{equation}
    ratio = \Delta m/\sigma_{\Delta m} \geq 4.5.
\end{equation}
The $ratio$ parameter acts to exclude noisy sources from our catalog of candidates, serving a role similar to a traditional signal-to-noise limit.  Third, each source must be extremely compact -- either unresolved or marginally resolved in our images -- to be selected.  This latter criterion is somewhat subjective, but is meant to exclude extended galaxies with emission from disk \ion{H}{2} regions.  This criterion also means that we do not ever select the AHA target galaxy located within each field as an H$\alpha$ Dot.

Our automated selection process also finds dozens of \ion{H}{2} regions in spiral disks and extended irregular galaxies.  In particular, the AHA target galaxy in each of our fields is often detected multiple times because of its disk \ion{H}{2} regions.   These \ion{H}{2} regions are identified during the visual examination of the computer-selected candidates and are cataloged separately.  In general, the distinction between the \ion{H}{2} regions and {\it bona fide} H$\alpha$ Dots is clear, as most disk \ion{H}{2} regions are located in the higher surface brightness portion of the stellar disk of their host galaxy.   In cases where there is substantial separation between the apparent outer edge of the disk and the line-emitting source, we include the object in the list of H$\alpha$ Dots.   We do not employ a rigorous criterion to select between the two options (\ion{H}{2} region vs. H$\alpha$ Dot), but in most cases the object must be at least 0.5 to 1.0 apparent disk radii beyond the outer edge of the galaxy to be included as an H$\alpha$ Dot.  As described below, several of these objects located at large radii are found through subsequent spectroscopy to be outlying \ion{H}{2} regions of the nearby galaxy. 

\section{Second List of H$\alpha$ Dot Candidates} 

The AHA data for seven observing runs carried out on the WIYN 0.9-m telescope were searched for H$\alpha$ Dots following the methods described above.   This represents a total of 354 fields with a total sky coverage of 22.72 deg$^2$.  As a result of these searches 119 H$\alpha$ Dots were detected.  The resulting surface density of H$\alpha$ Dots is 5.24 deg$^{-2}$, virtually identical to the density of Dots found in \citet{hadot1} of 5.22 deg$^{-2}$.

\begin{figure*}[ht]
\centering
\epsscale{0.9}
\plotone{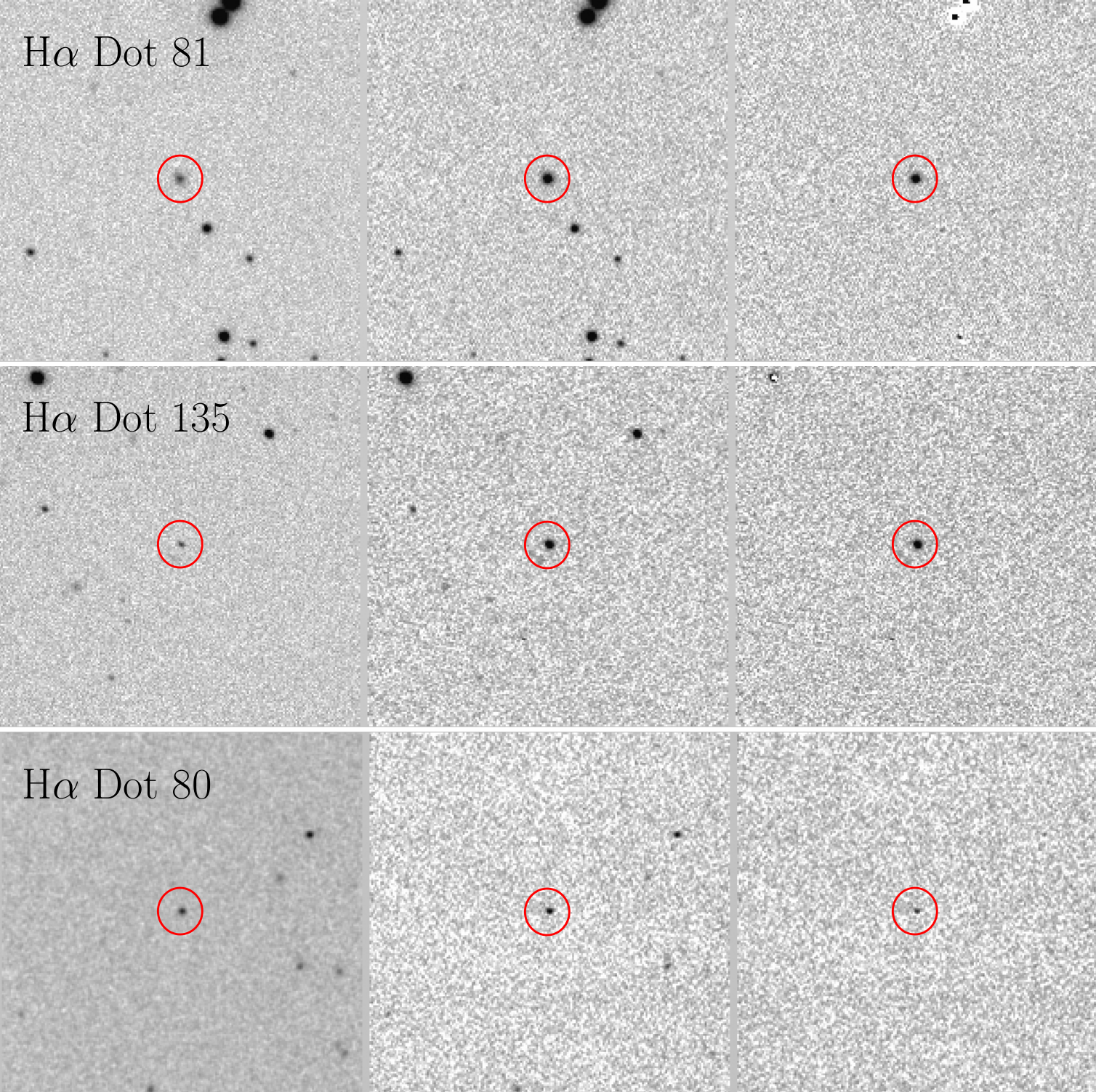}
\caption{\footnotesize Example images of H$\alpha$ Dots cataloged in this study.  In all three examples, the leftmost image is a cut-out of the R-band continuum image, the central image is taken through the relevant narrow-band filter, and the rightmost image is the continuum-subtracted image.  Each image section shows a 120 $\times$ 120 arcsec FOV.  {\bf Top:} H$\alpha$ Dot 81 is an H$\alpha$-detected dwarf star-forming galaxy with R = 18.85, M$_R$ = $-$15.5, and z = 0.0153.  {\bf Middle:} H$\alpha$ Dot 135 is an [\ion{O}{3}]-detected Green Pea-like galaxy with R = 20.17, M$_R$ = $-$21.3, and z = 0.3294.  {\bf Bottom:} H$\alpha$ Dot 80 is a QSO detected by its Ly$\alpha$ line.  It has R = 19.82, M$_R$ = $-$28.5, and z = 4.4937.  All three objects shown here have spectra displayed in \S 4.}
\label{fig:examples}
\end{figure*}

Table~\ref{tab:prop} presents our second catalog of H$\alpha$ Dot candidates.  The first column in the table is the H$\alpha$ Dot number.   Following the convention established in \citet{hadot1}, the H$\alpha$ Dot candidates are numbered sequentially in right ascension (RA) order for all objects found from the search of the fields from a given AHA observing run.  For example, the search of the data obtained in March 2008, the first of the AHA observing runs considered in the current catalog, yielded 15 dot candidates.  These were ordered by RA and numbered 62 -- 76, since the previous list ended with H$\alpha$ Dot 61.  Following the dot number, RA and Dec (J2000) are given in columns 2 and 3.  The coordinates listed are typically those measured by the Sloan Digital Sky Survey \citep{sdss,sdss14}, as we found that the SDSS astrometry was typically more reliable than our own.

The selection parameters $\Delta$m and $ratio$ are listed in columns 4 and 5.  These are defined in the previous section.  Column 6 lists the measured R-band magnitude of each source as obtained from our continuum images.  Comparison between our photometry and the r-band magnitudes from SDSS generally show good agreement.  Column 7 presents the emission-line flux measured from the continuum-subtracted narrow-band images.  The narrow-band fluxes are calibrated using spectrophotometric standard stars obtained nightly as part of the AHA survey. We note that these emission-line fluxes are {\it not} H$\alpha$ fluxes in all cases.  As discussed in \S 4 and \S 5, for many of the objects in our survey the emission line detected is a line other than H$\alpha$.

The final three columns in Table~\ref{tab:prop} are derived properties of the H$\alpha$ Dots.  We present the R-band absolute magnitude in column 8, the H$\alpha$ luminosity in column 9, and the H$\alpha$-based star-formation rate (SFR) in column 10.  All three quantities make use of the spectroscopic redshifts presented in \S 4.  We discuss these derived properties in detail in \S5.4.

\begin{figure*}[ht]
\centering
\includegraphics[width=0.49\linewidth]{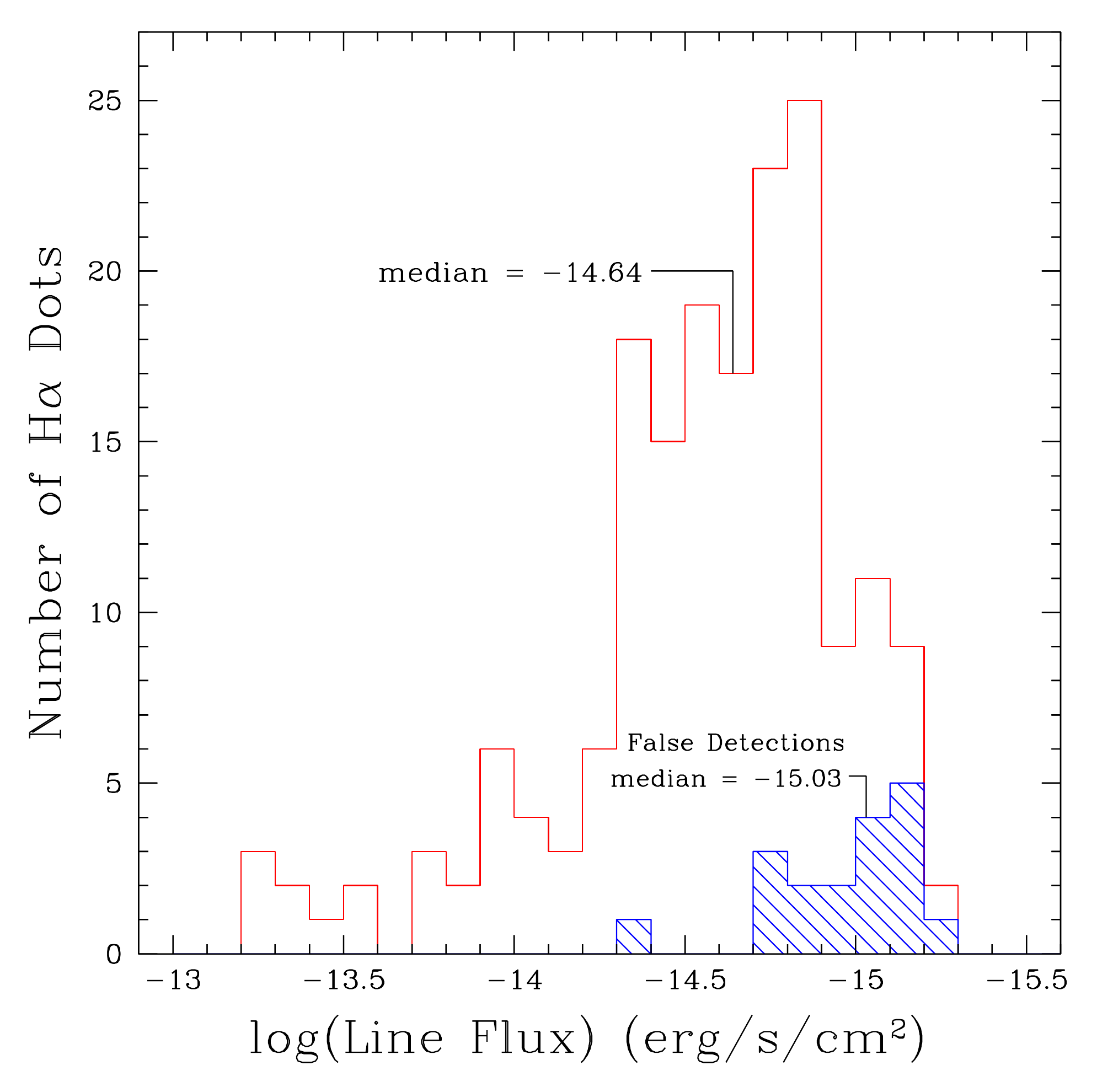}\quad\includegraphics[width=0.49\linewidth]{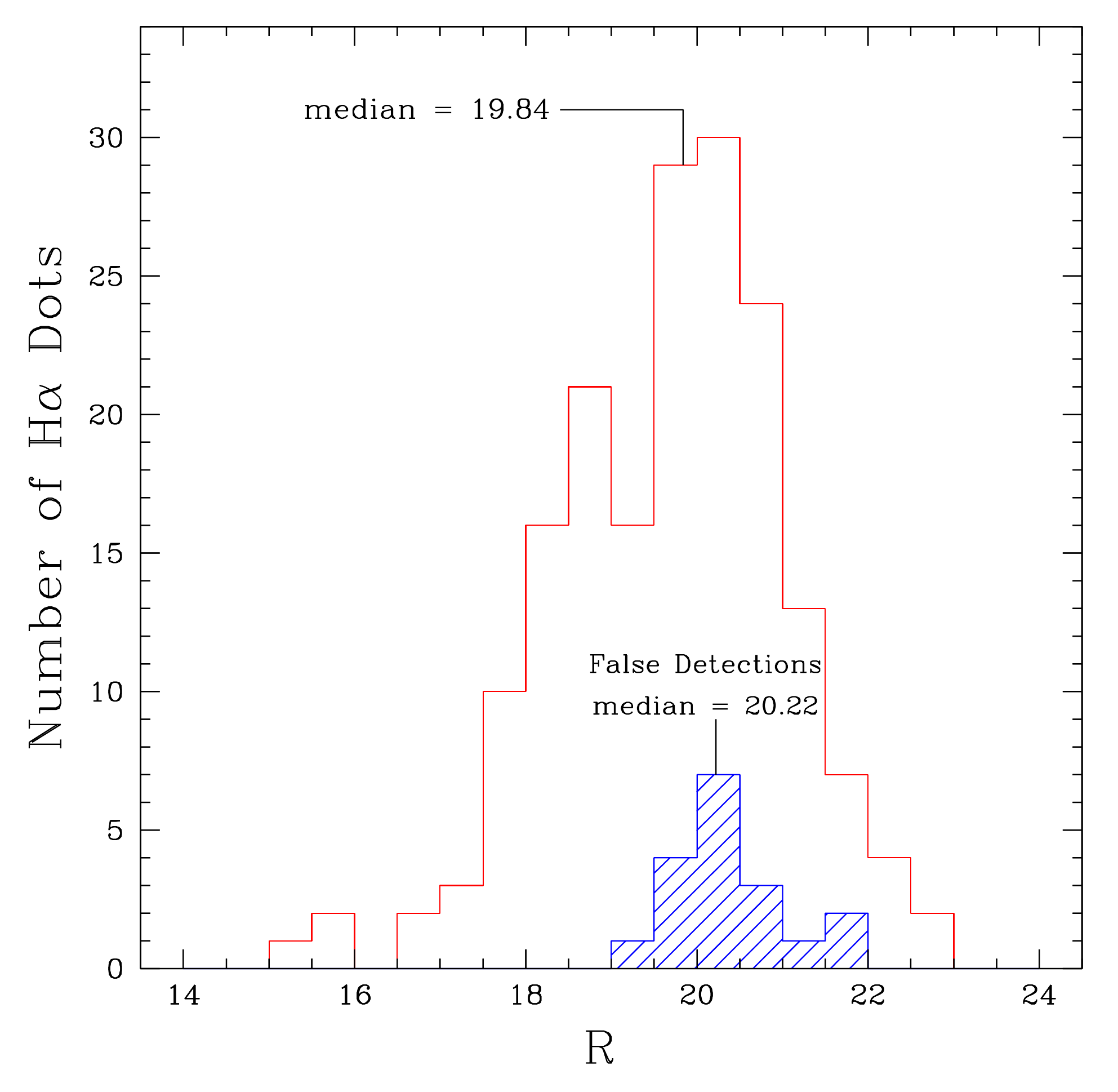}
\caption{\footnotesize (left) Composite emission-line flux distribution for the first two lists of  H$\alpha$ Dots (N=180).  The line flux is measured directly from our narrow-band images.   The specific emission line being measured is only revealed through spectroscopic follow-up.  The blue cross-hatched histogram shows the line flux distribution of the 18 false detections.   The latter distribution is heavily weighted toward the faintest fluxes detected by our survey.  (right) Composite R-band magnitude distribution for the first two lists of H$\alpha$ Dots (N=180).   The median apparent magnitude of R = 19.84 is indicated.  The blue cross-hatched histogram shows the magnitude distribution for the 18 false detections; the majority of these objects are included in the fainter half of the overall sample.}
\label{fig:r_hist}
\end{figure*}

Images of three example H$\alpha$ Dots are shown in Figure~\ref{fig:examples}.  In each example, the leftmost image is a cut-out of our R-band continuum image, the middle image is the same field imaged with the relevant narrow-band filter, and the rightmost image shows the continuum-subtracted image.  All image sections are 2.0 $\times$ 2.0 arcminutes square.  These three objects were selected because they illustrate the three principal line-selection groups discussed in \S 4.  The top row shows the H$\alpha$-detected star-forming galaxy H$\alpha$ Dot 81, the middle row displays the [\ion{O}{3}]-detected galaxy H$\alpha$ Dot 135, while the bottom row presents the Ly$\alpha$-detected QSO H$\alpha$ Dot 80.  All three of these objects also have spectra that are shown in Figures~\ref{fig:speclz}-\ref{fig:specqso}.  While all three objects clearly show excess emission in the continuum-subtracted images (right-hand column), it is not possible to ascertain the galaxy activity type or which emission line was present in the survey narrow-band filter by examination of the images alone.

Figure~\ref{fig:r_hist} presents histograms of the two principal observables from the imaging survey data: emission-line flux and R-band apparent magnitude.  In both plots we show the cumulative distributions from both of the survey lists.  The red-lined histogram shows the measured values for all 180 H$\alpha$ Dots, while the blue cross-hatched histograms show the same parameters for the 18 false detections.  The latter were identified after completion of the follow-up spectroscopy described in \S 4.

The distribution of emission-line flux is shown in the left-hand figure, and exhibits a median value of 2.3 $\times$ 10$^{-15}$ erg/s/cm$^2$.  The flux distribution rises until $\sim$1.3 $\times$ 10$^{-15}$ erg/s/cm$^2$, beyond which it drops off precipitously.  This flux level can be taken as an {\it approximate} line flux completeness limit for the sample.  It should be stressed that the detection of an emission-line source in a narrow-band survey such as this depends on a number of factors: line flux, line equivalent width (e.g., the contrast between the line and the underlying stellar continuum), and redshift.  Despite the complicated selection function, the flux value associated with the peak of the histogram can still serve as a first-pass estimate of the survey depth.  The faintest ``real" objects detected have line fluxes of $\sim$6 $\times$ 10$^{-16}$ erg/s/cm$^2$.  Not surprisingly, the false detections are nearly all found among the lowest flux objects in the survey.  Only one has a measured line flux within the brighter half of the sample.

The R-band apparent magnitude distribution is shown in the right half of  Figure~\ref{fig:r_hist}.  The overall brightness range of the sample is quite large, extending from R = 15 to R = 23, with a median value of R = 19.84.  The false detections once again tend to be located among the fainter half of the overall distribution.   However, this tendency is not nearly as strong as seen in the line flux plot, a clear indication that the sample is {\bf not} a magnitude-limited one.  Fully 23 out of the 26 objects with R $>$ 21.0 are {\it bona fide} emission-line objects, including all six with R $>$ 22.0.  We consider this to be an impressive result considering that the survey was carried out on a 0.9-m telescope.

Due to the overlap of some of the AHA survey fields, several H$\alpha$ Dots have been detected more than once with images taken on separate observing runs.  This includes five objects in the current survey list that were previously cataloged in \citet{hadot1}, plus two additional sources that were found in data obtained on different observing runs covered by the current searches.  Since these duplicate detections represent discoveries made from independent data, we decided to retain them in our catalog for completeness.   Hence, our current catalog includes only 112 unique H$\alpha$ Dots.  All duplicate H$\alpha$ Dots are labeled as such in Table~\ref{tab:prop}.  In the subsequent discussions and figures these duplicate sources will appear only once.

\section{Follow-up Spectroscopy of the New H$\alpha$ Dot Candidates} 

\begin{deluxetable*}{ccccccccccc}
\tabletypesize{\footnotesize}
\tablewidth{0pt}
\tablecaption{H$\alpha$ Dots: Measured Spectral Properties \label{tab:spec}}

\tablehead{
 \colhead{H$\alpha$ Dot \#} &  \colhead{ELG Type} &  \colhead{z} &  \colhead{c$_{H\beta}$} &  \colhead{EW H$\alpha$} & \colhead{EW [\ion{O}{3}]} & \colhead{[\ion{N}{2}]/H$\alpha$} & \colhead{[\ion{O}{3}]/H$\beta$} &  \colhead{[\ion{O}{2}]/H$\beta$} &  \colhead{[\ion{S}{2}]/H$\alpha$}  &  \colhead{log(O/H)+12} \\
 (1)  & (2)  & (3)  & (4)  & (5)  & (6)  & (7)  & (8)  & (9)  & (10) & (11)
}
 \startdata
   62 & SFG &  0.0163 &  -0.02 &   64.4 &   64.2 &  -1.51 &   0.57 &  -- &  -0.86 &   8.03 \\
   63 & \ion{H}{2} &  0.0169 &   0.40 &   85.9 &   48.4 &  -0.71 &   0.45 &  -- &  -0.55 &   8.55 \\
   64 & NotELG &  -- &  -- &   -- &    -- &  -- &  -- &  -- &  -- &  -- \\
   65 & SFG &  0.3243 &   0.24 &  369.2 &  360.7 &  -1.26 &   0.64 &   0.37 &  -0.72 &   8.15 \\
   66 & SFG &  0.0098 &  -0.17 &   40.8 &   23.0 &  -1.05 &   0.32 &  -- &  -- &  8.44 \\
   67 & SFG &  0.0145 &   0.74 &   37.4 &    3.9 &  -0.47 &  -0.40 &  -- &  -0.50 &   8.97 \\
   68 & NotELG &  -- &  -- &   -- &    -- &  -- &  -- &  -- &  -- &  -- \\
   69 & SFG &  0.0157 &   0.54 &   22.1 &    5.9 &  -1.22 &   0.20 &  -- &  -0.53 &  -- \\
   70 & Sy2 &  0.3239 &   0.69 &   63.5 &  126.8 &  -0.19 &   0.92 &   0.57 &  -- &  -- \\
   71 & Sy2 &  0.3313 &  -- &   13.8 &   19.1 &  -0.46 &   1.41 &   1.12 &  -- &  -- \\
\\
   72 & SFG &  0.3325 &   0.47 &   -- &  125.8 &  -- &   0.38 &   0.57 &  -- &  -- \\
   73 & SFG &  0.3173 &   0.31 &  123.2 &   69.3 &  -1.25 &   0.45 &   0.46 &  -0.48 &   8.27 \\
   74 & \ion{H}{2} &  0.0128 &   0.73 & 1319.0 &  603.1 &  -0.96 &   0.43 &  -- &  -0.81 &   8.44 \\
   75 & QSO &  4.4637 &  -- &   -- &    -- &  -- &  -- &  -- &  -- &  -- \\
   76 & SFG &  0.3299 &   0.53 &  224.0 &  220.4 &  -1.26 &   0.54 &   0.55 &  -- &   8.21 \\
   77 & SFG &  0.0125 &   0.20 &   46.8 &   21.9 &  -0.82 &   0.25 &  -- &  -0.49 &   8.59 \\
   78 & SFG &  0.0170 &   0.09 & 1706.0 &  646.6 &  -1.67 &   0.55 &  -- &  -1.21 &   7.94 \\
   79 & SFG &  0.0168 &   0.05 &  177.4 &  131.9 &  -1.44 &   0.51 &  -- &  -0.77 &   8.12 \\
   80 & QSO &  4.4937 &  -- &   -- &    -- &  -- &  -- &  -- &  -- &  -- \\
   81 & SFG &  0.0153 &   0.18 &  364.8 &  562.7 &  -1.64 &   0.77 &  -- &  -1.15 &   7.80 \\
 \enddata
 \tablecomments{Table 2 is published in its entirety in the machine-readable format.  A portion is shown here for guidance regarding its form and content.}

\end{deluxetable*} 

Once an H$\alpha$ Dot candidate is detected in the narrow-band images, it is placed into a queue for follow-up spectroscopy.  Until a spectrum is obtained for an H$\alpha$ Dot candidate, it remains just that: a candidate.   Furthermore, without a follow-up spectrum, we have no idea which emission line was responsible for the detection, or what class of activity the H$\alpha$ Dot belongs to (e.g., star forming or AGN).  Hence, the acquisition of a follow-up spectrum is essential for both verifying and understanding the nature of each candidate.

For all of the H$\alpha$ Dots in the current study, follow-up ``first-look" spectra were obtained using the Hobby-Eberly 9.2-m telescope\footnote{Based on observations obtained with the Hobby-Eberly Telescope, which is a joint project of the University of Texas at Austin, the Pennsylvania State University, Stanford University, Ludwig-Maximilians-Universit\"at M\"unchen, and Georg-August-Universit\"at G\"ottingen.} (HET) located at McDonald Observatory.   All observations were carried out in queue mode by the resident HET observers.  As H$\alpha$ Dots were cataloged they were forwarded to the HET queue to await observation, usually with a one year time lag.  All of these ``first-look" spectra were taken with short exposure times of 10 minutes each.

All spectral data were obtained with the Marcario low-resolution spectrograph \citep[LRS;][]{hill1998}.  Initially, each candidate was observed in the wavelength region 4350--7250 \AA\ using LRS grating G2.  The G2 spectra have a reciprocal disperson of 2.00 \AA\ pixel$^{-1}$ and a pixel scale of 0.47 arecsec along the slit (for the default case of 2 $\times$ 2 pixel binning).  These data provided coverage of the spectral region containing the emission line detected in the narrow-band filter images and allowed for the determination of the redshift of each source.  In the cases where the H$\alpha$ Dot was detected via the H$\alpha$ line, this was the only spectrum obtained.  Galaxies detected by the [\ion{O}{3}]$\lambda$5007 line were left in the queue to be observed with the G3 grating.  The purpose of the G3 spectra was to observe redshifted H$\alpha$ and the [\ion{N}{2}] $\lambda\lambda$6383,6548 doublet.  The G3 grating delivered spectra that covered the wavelength range 6300--9100 \AA\ at 1.91 \AA\ pixel$^{-1}$.  [\ion{O}{2}]-detected galaxies were also observed with the G3 grating, giving us access to their [\ion{O}{3}] and H$\beta$ lines.  All spectra were obtained using a 2~arcsecond wide slit.

Data processing followed standard practice, and all data reduction was carried using IRAF.  
The spectral images had their bias levels removed using overscan fitting, after which an averaged bias image was subtracted to remove any two-dimensional structure.  Flat-field frames were combined into a single image, normalized, and applied to the science images.  Particle events on the detector were eliminated using LACOS\_SPEC \citep{vandokkum2001}.  Next the spectra were extracted to a one-dimensional format and sky subtracted.  Wavelength calibrations were provided using spectra of Ne and Cd lamps, and a single spectrophotometric standard star observed nightly was used to perform a relative flux calibration.  The emission lines present in the fully reduced spectra were measured using the IRAF routine SPLOT. 

All of the H$\alpha$ Dots listed in Table~\ref{tab:prop} have been observed spectroscopically.  We tabulate relevant information derived from our spectra in Table~\ref{tab:spec}.  Column 1 lists the H$\alpha$ Dot number while column 2 specifies the activity type of each object: SFG = star-forming galaxy, \ion{H}{2} = outlying/isolated \ion{H}{2} region, Sy1 = Seyfert 1 galaxy, Sy2 = Seyfert 2 galaxy, QSO = quasi-stellar object, and NotELG = a false detection whose spectrum displays no emission lines.  The criteria used to place each H$\alpha$ Dot into the appropriate activity class are discussed in \S 5.2 and \S 5.3.   Column 3 gives the observed redshift z, and column 4 lists the decimal reddening coefficient c$_{H\beta}$, nearly always derived from the observed H$\alpha$/H$\beta$ ratio.   In a few cases  of higher-redshift systems where H$\alpha$ is not observed we use the H$\gamma$/H$\beta$ ratio instead.   Negetive values of c$_{H\beta}$ are treated as zero when correcting the emission-line ratios for reddening.  Columns 6 and 7 give the measured equivalent widths of the H$\alpha$ and [\ion{O}{3}]$\lambda$5007 lines, respectively (in \AA).  The next four columns (7 - 10) list the logarithms of the specified ratios of emission-line fluxes.  The line ratios have all been corrected for reddening using the value of c$_{H\beta}$ given in column 4.  Because the wavelength coverage of our spectra is fixed, but the redshifts of the objects vary over a huge range, many of the line ratios indicated are not available for a given object.  Finally, column 11 lists the derived oxygen abundance, given in the standard nebular scale of log(O/H) + 12.  A description of how the abundances are derived is given in \S 5.

\begin{figure}[t]
\epsscale{1.2}
\plotone{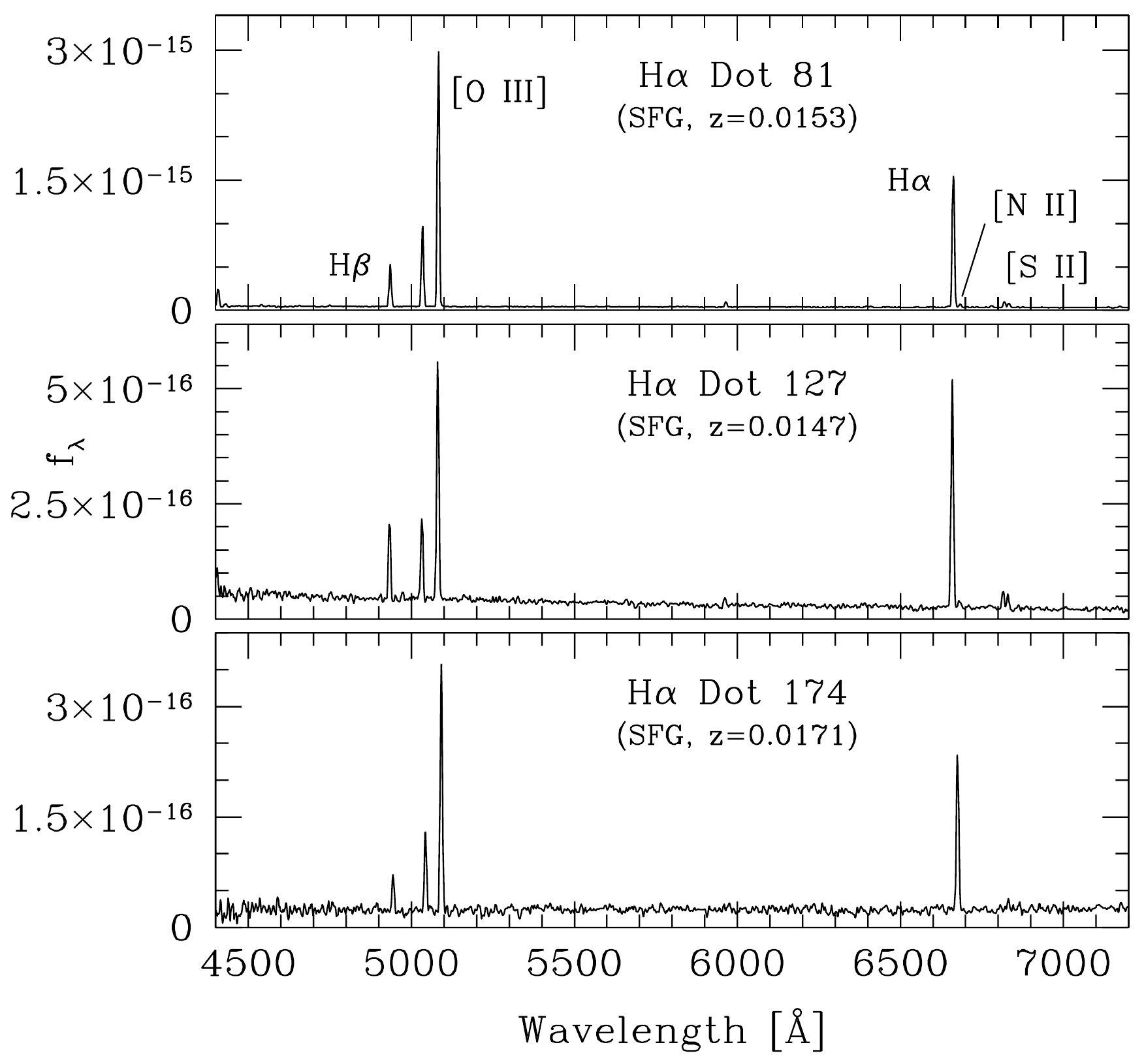}
\caption{\footnotesize Example spectra of H$\alpha$ Dot galaxies.  The three spectra included in this figure are all low redshift H$\alpha$-detected star-forming systems.  They are all dwarf galaxies and have low metallicities.  }
\label{fig:speclz}
\end{figure}

\begin{figure}[t]
\epsscale{1.2}
\plotone{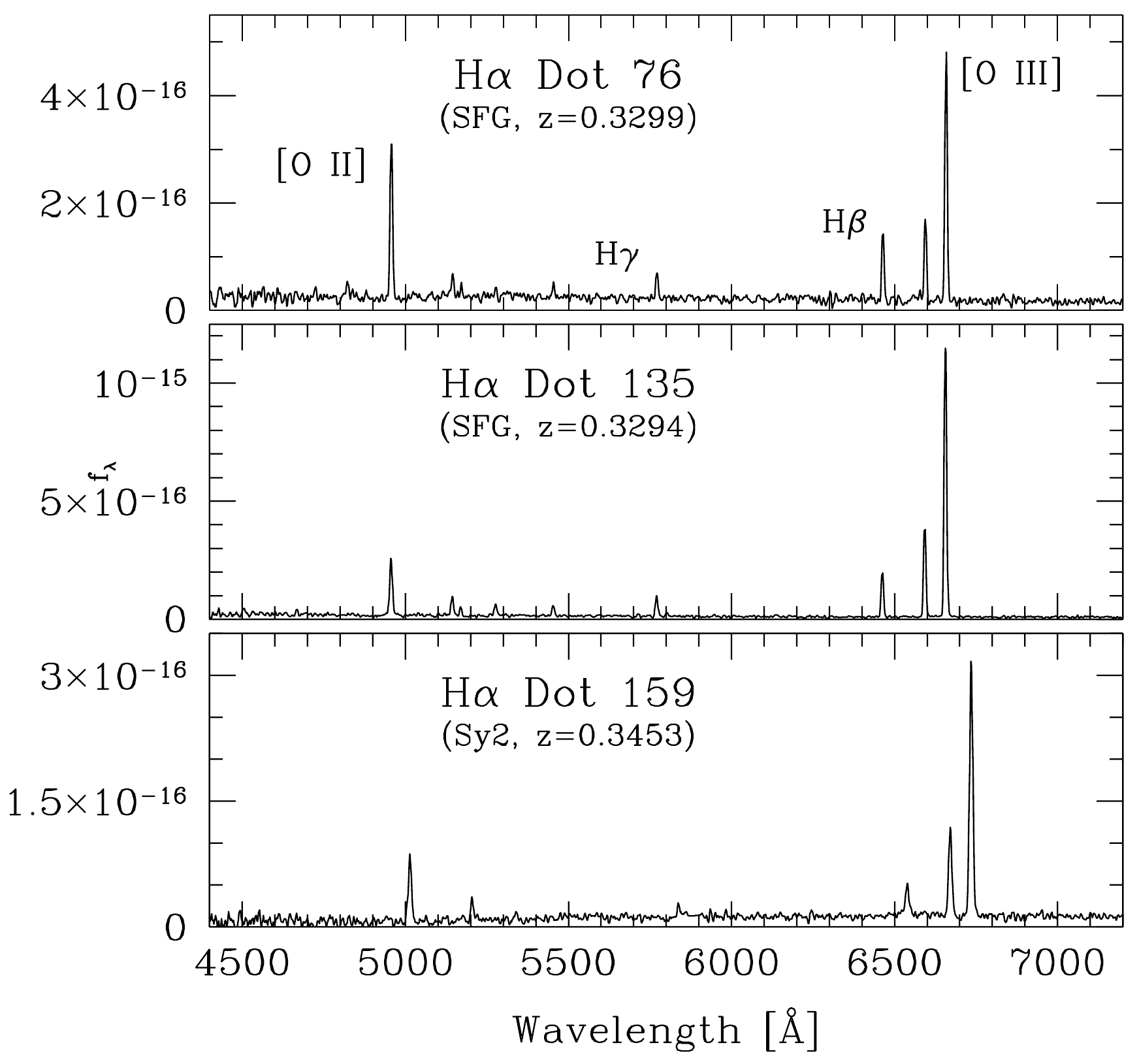}
\caption{\footnotesize Example spectra of three [O~III]-detected H$\alpha$ Dots.  In all three cases, we possess additional spectra that reach further to the red and include the H$\alpha$ and [N~II] lines.  This allows us to unambiguously classify the galaxies as either star-forming or AGN.  The upper two spectra illustrate SFGs, with H$\alpha$ Dot 135 exhibiting properties similar to the  Green Pea galaxies, while the bottom spectrum is that of a Seyfert 2 galaxy.}
\label{fig:spechz}
\end{figure}

\begin{figure}[ht]
\epsscale{1.2}
\plotone{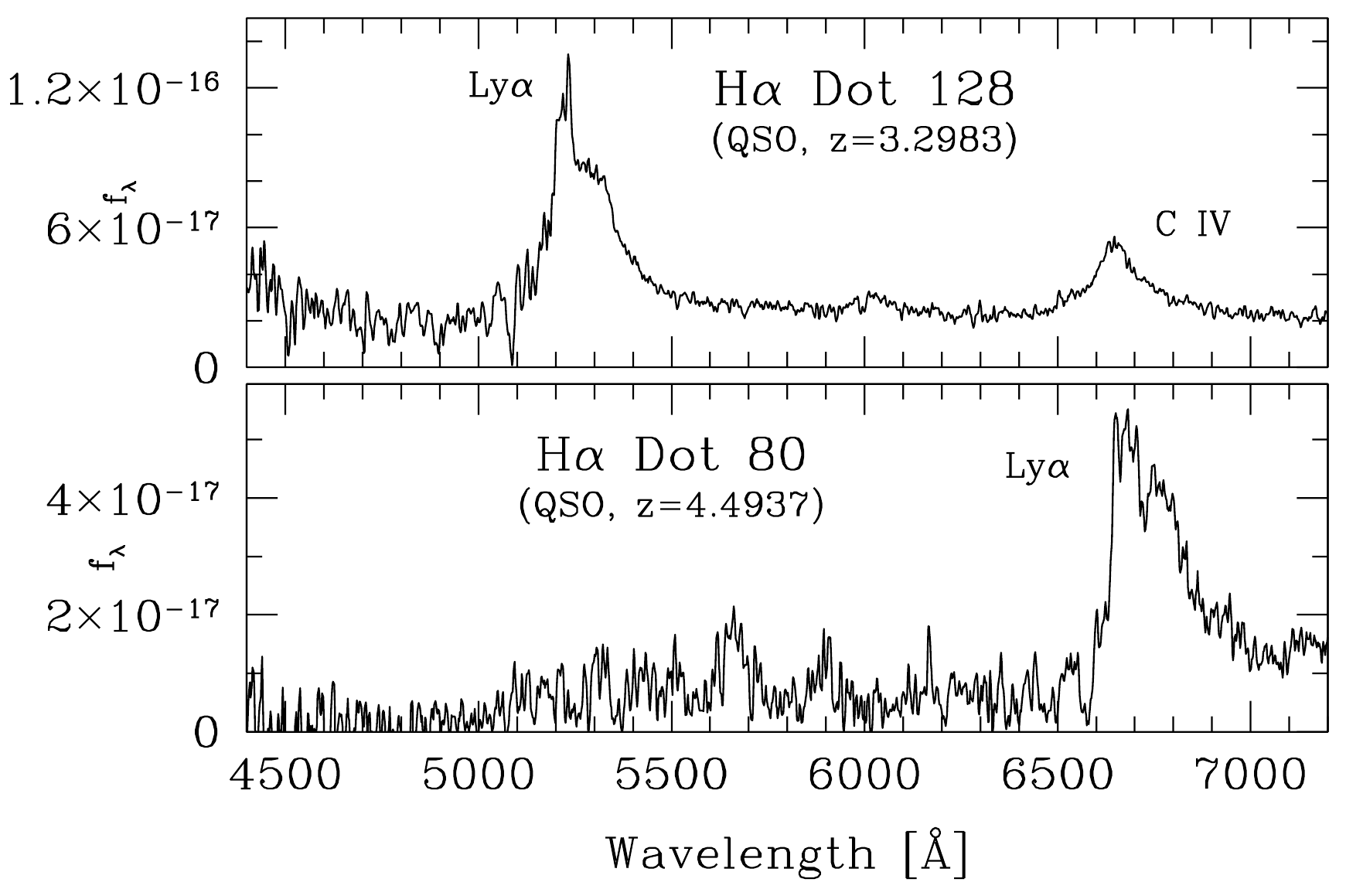}
\caption{\footnotesize Spectra of two QSOs detected in the H$\alpha$ Dot survey.  The upper spectrum represents a z = 3.2983 QSO that was detected via its strong C IV $\lambda$1549 line; Ly$\alpha$ is also present.  The lower spectrum illustrates a Ly$\alpha$-detected QSO with z = 4.4937.}
\label{fig:specqso}
\end{figure}

Plots displaying example spectra are shown in Figures~\ref{fig:speclz} - \ref{fig:specqso}.   We separate the spectra into the three principal line-selection categories present in the survey: H$\alpha$ detection, [\ion{O}{3}] detection, and detection via a UV emission line.  The spectral coverage of the narrow-band filters used for the survey was between 6595 and 6734 \AA, so the specific emission line responsible for the detection of each H$\alpha$ Dot is usually immediately apparent from its spectrum.

Figure~\ref{fig:speclz} illustrates three H$\alpha$-detected sources.  Objects detected via the H$\alpha$ line are the most common type of H$\alpha$ Dot.  In all three cases, the spectra show strong [\ion{O}{3}]$\lambda\lambda$4959,5007 lines and weak [\ion{N}{2}]$\lambda\lambda$6548,6583 lines, indicative of metal-poor galaxies.  Unfortunately, the spectral range provided by the HET LRS set-up used does not cover the important [\ion{O}{2}]$\lambda\lambda$3726,3729 doublet for the low redshifts of the H$\alpha$-detected H$\alpha$ Dots.  Two of the galaxies illustrated -- H$\alpha$ Dot 81 and 174 -- are blue compact dwarfs (BCDs), with R-band absolute magnitudes of $-$15.5 and $-$15.2, respectively.  Both exhibit the very compact morphology characteristic of BCDs.  The third, H$\alpha$ Dot 127, represents a luminous \ion{H}{2} region located on the edge of a very low surface brightness disk.   The disk was not detected in our short R-band survey images, which is why it was classified as an H$\alpha$ Dot.  H$\alpha$ Dot 127 is still a very dwarfy system (M$_R$ = $-$16.0).  A more complete evaluation of the properties of the H$\alpha$-detected objects is presented in the follow section.

Detections via the [\ion{O}{3}]$\lambda$5007 line account for the second largest sub-sample of objects cataloged in the H$\alpha$ Dot survey.  Figure~\ref{fig:spechz} shows the spectra of three [\ion{O}{3}]-detected objects.  The top spectrum is of H$\alpha$ Dot 76, a star-forming galaxy with M$_R$ = $-$20.5.  It appears to be part of a merger system.  The spectrum of H$\alpha$ Dot 135 (middle) exhibits the characteristics of the Green Pea galaxies \citep[e.g.,][]{gp, brunker2020}, and in fact looks like a Green Pea in the SDSS color images.  With an [\ion{O}{3}]$\lambda$5007 equivalent width of $\sim$1000 \AA\ and a very compact appearance (unresolved in our images) it certainly appears to be a true Green Pea.  Finally, the lower spectrum in Figure~\ref{fig:spechz} is that of the Seyfert 2 galaxy H$\alpha$ Dot 159.  The similarities between this spectrum and the one directly above it in Figure~\ref{fig:spechz} illustrate the importance of having the G3 spectra that cover the H$\alpha$ plus [\ion{N}{2}]$\lambda\lambda$6548,6583 lines.  It is the H$\alpha$/[\ion{N}{2}] ratio that clearly delineates this galaxy as a Seyfert 2 (see \S5.2 and \S5.3 for details of our classification criteria).   H$\alpha$ Dot 159 has an R-band absolute magnitude of $-$21.1.

Figure~\ref{fig:specqso} shows two spectra of higher-redshift H$\alpha$ Dots that were detected by one of the strong UV lines typically seen in the spectra of QSOs.  The top spectrum is that of H$\alpha$ Dot 128, a QSO detected via its \ion{C}{4}$\lambda$1549 line at z = 3.30, while the lower spectrum is that of the Ly$\alpha$-detected QSO H$\alpha$ Dot 80 (z = 4.49).  Even though the UV-line-detected component of the survey is small ($\sim$7\%), we consider it an impressive accomplishment of the narrow-band selection technique that we are sensitive to z = 4 QSOs using an 0.9-m telescope.

Nine of the objects in our second catalog of H$\alpha$ Dots did not exhibit any emission lines in their follow-up spectra, and are labeled as false detections in Table~\ref{tab:spec}.  This represents a false positive rate of 7.6\%, which is roughly half of the value from the first survey list.  We used the results from the first survey to improve upon our selection method, resulting in this dramatic drop in false positives.  In most cases, the false detections are objects with line fluxes close to our sensitivity limit (Figure~\ref{fig:r_hist}), although in a few instances they were caused by image artifacts that escaped attention during our visual checks of the candidates.  Of the nine false detections in the current catalog, four are galactic stars and five are galaxies.

\section{Discussion} 

\subsection{Redshifts}

The narrow-band selection technique employed in the H$\alpha$ Dots survey naturally leads to the detection of discrete classes of emission-line objects in well-defined redshift ranges.  In this subsection we examine the redshift distribution of the H$\alpha$ Dots, while in the following subsection we utilize the spectroscopic data to classify each object into one of a handful of specific groups.  Throughout the remainder of this paper we combine the H$\alpha$ Dots cataloged in both the first survey list \citep{hadot1} as well as those discovered in the current work to take advantage of the larger sample of objects.   Our evaluation suggests that there are no fundamental differences between the properties of the objects from the two catalogs.

\begin{figure}[t]
\epsscale{1.2}
\plotone{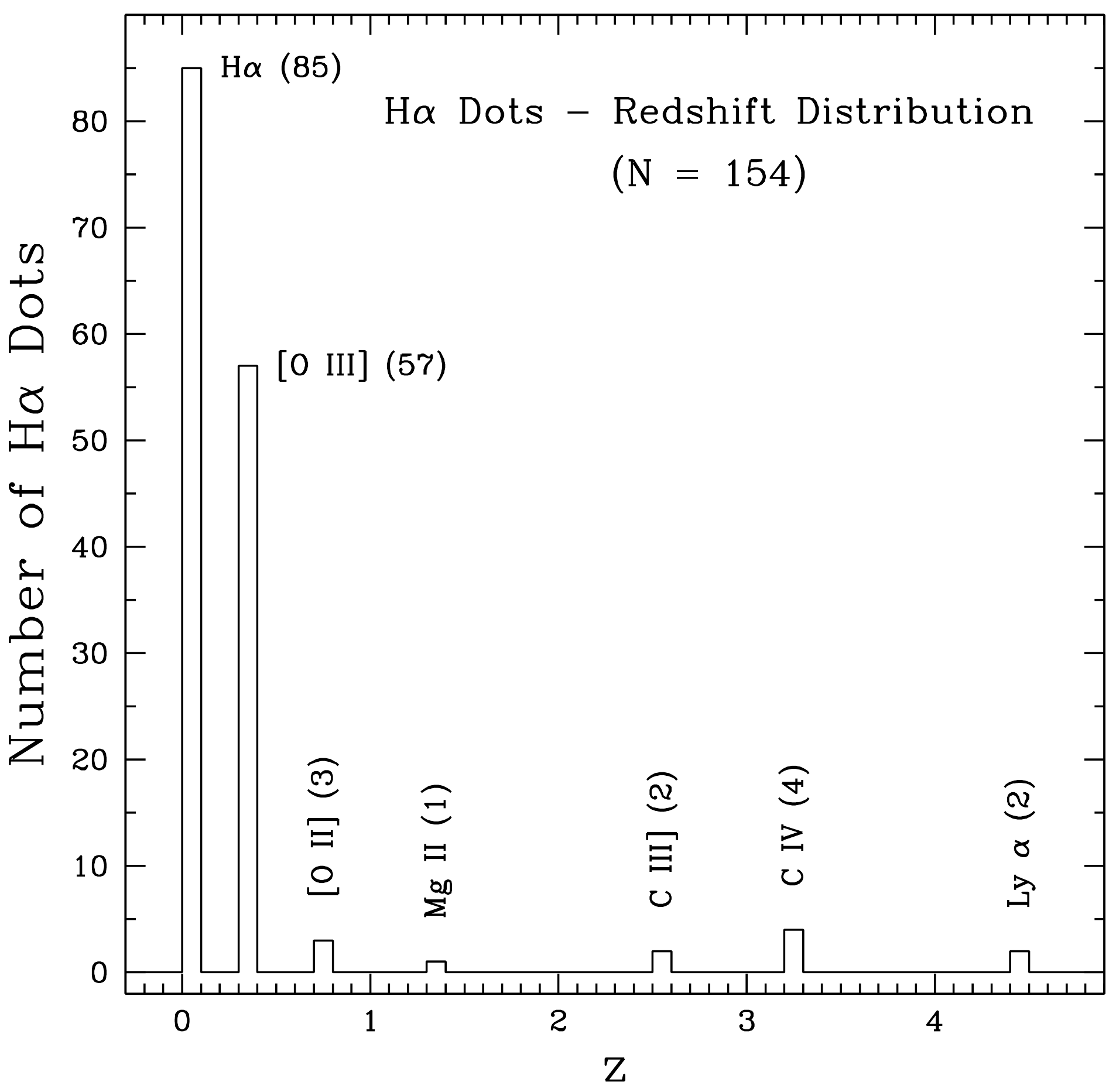}
\caption{\footnotesize Redshift distribution of the H$\alpha$ Dots from the first two survey lists (N=154 unique objects).  Each peak in the plot represents a specific emission line that has shifted into the bandpass of the filters used from the ALFALFA H$\alpha$ survey.}
\label{fig:redshifts}
\end{figure}

Figure \ref{fig:redshifts} displays the observed redshift distribution for the H$\alpha$ Dots from both survey lists.  There are 154 unique objects included rather than 180.   The difference is accounted for by the 18 false detections in the first two lists and the 8 duplicate objects found in the analysis of independent survey images.  The redshift distribution is displayed with a very coarse sampling on purpose, so that the height of each bin represents the number of objects detected via a specific emission line.  

By far the most common lines used in the discovery of the H$\alpha$ Dots are H$\alpha$ and [\ion{O}{3}]$\lambda$5007.  The H$\alpha$ line accounts for 85 out of the 154 unique ELGs cataloged in the first two lists (55\%).  The H$\alpha$-detected sources fall within the redshift range 0.0056 - 0.0243 (velocity range 1690 -- 7430 km/s).  As discussed below, they are predominantly low-luminosity systems.  The [\ion{O}{3}]-detected H$\alpha$ Dots represent an additional 57 sources (37\% of the sample).  In order to be detected via the [\ion{O}{3}] line using our survey filters, objects need to be in the redshift range 0.316 -- 0.349 (distances of 1660 -- 1875 Mpc).   This in turn requires them to be fairly luminous systems in order to be detected. 

The remaining 12 H$\alpha$ Dots are higher redshift systems that were detected via one of several UV emission lines.  Three sources were detected by the [\ion{O}{2}]$\lambda$3727 doublet, and have redshifts between 0.78 and 0.80.  Based on our G3 spectra of these galaxies, we classify two as AGN and one as star forming (see \S5.2 and \S5.3 for classification details).  The remaining nine high-redshift H$\alpha$ Dots are all broad-lined QSOs.   These include one object detected via its \ion{Mg}{2} ($\lambda$2798) doublet (z = 1.35), two found by the semi-forbidden \ion{C}{3}]$\lambda$1909 line (z = 2.49 -- 2.55), four detected from strong \ion{C}{4} 
($\lambda$1549) at z between 3.28 and 3.30, and two via Ly$\alpha$ emission ($\lambda$1215) with redshifts of 4.46 and 4.49 (see Figure~\ref{fig:specqso}).  While the higher redshift sources make up only 8\% of the overall sample, they represent interesting sources and emphasize the power of the narrow-band detection method for detecting faint, high-redshift objects.

\subsection{Spectroscopic Diagnostics}

\begin{figure}[t]
\epsscale{1.2}
\plotone{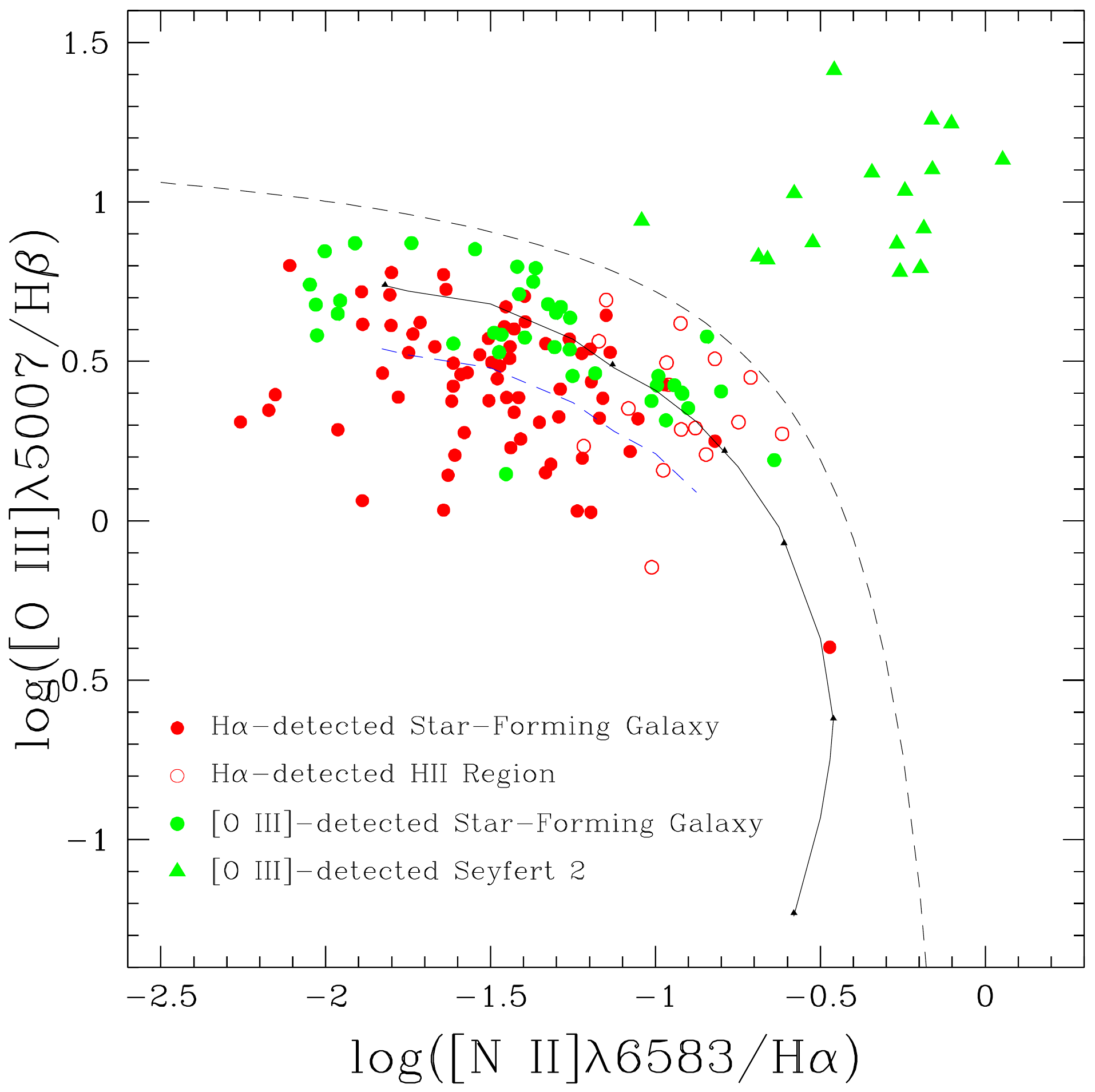}
\caption{\footnotesize Emission-line diagnostic diagram for the full sample of H$\alpha$ Dots.   This figure includes 138 out of the 154 unique, {\it bona fide} Dots, excluding the broad-lined AGN (QSOs and Seyfert 1s) as well as four galaxies that lack detection of both line ratios plotted.   The solid line shows a locus of theoretical model points taken from \citet{DE86}, where the assumed metallicity of the gas decreases along the model curve from the lower right (2$\times$ solar abundance) to the upper left (0.25$\times$ solar abundance).  The black dashed line shows the empirical boundary between stellar and non-stellar photo-ionized objects \citep{kauff03}.  The meaning of the blue dashed line is discussed in \S 5.4.  The significance of the locations of each of the four classes of objects displayed are discussed in the text.}
\label{fig:diag}
\end{figure}

We can learn much about the nature of our Dots by considering the information presented in their emission-line spectra.  In particular, we can utilize an emission-line ratio diagnostic diagram \citep[e.g.,][hereafter BPT diagram]{bpt,vo1987} using the measured strengths of key emission lines to help distinguish between different types of objects.  Figure \ref{fig:diag} displays such a BPT diagram for the H$\alpha$ Dots, plotting the logarithm of the [\ion{O}{3}]$\lambda$5007/H$\beta$ ratio against the logarithm of the [\ion{N}{2}]$\lambda$6583/H$\alpha$ ratio.

We display the objects detected with the H$\alpha$ line in red and objects detected via [\ion{O}{3}]$\lambda$5007 in green in Figure \ref{fig:diag}.   We further subdivide both groups.  For the H$\alpha$ sources, we distinguish between objects that are distinct star-forming galaxies (solid red dots) and those that are outlying \ion{H}{2} regions near large disk galaxies (red circles).   The latter are usually located well outside the apparent optical disk of the host galaxy (see below).  The outlying \ion{H}{2} regions as a class tend to be located among the more metal-rich objects located in Figure \ref{fig:diag} (toward higher [\ion{N}{2}]$\lambda$6583/H$\alpha$ ratios); as we discuss below this is not unexpected.  The majority of the star-forming galaxies exhibit lower values of [\ion{N}{2}]$\lambda$6583/H$\alpha$, indicative of metal-poor systems.

It is important to stress that the inclusion of an emission-line object in the H$\alpha$ Dots survey requires that the source be either extremely compact or unresolved at the resolution of our ground-based images (typical image quality of 1-2\arcsec ).   This in turn impacts dramatically the make-up of our low-redshift H$\alpha$-detected sources.   The H$\alpha$ Dot survey does not catalog extended objects with large angular extents.  Hence, luminous spiral disks with multiple \ion{H}{2} regions are excluded from our lists, as are extended galaxies with strong nuclear starbursts.  The natural outcome of this aspect of our selection function is that the H$\alpha$-detected component of the survey are typically low-luminosity star-forming dwarfs.  This in turn explains the tendency for the H$\alpha$-detected objects to be located in the upper left portion of the star-forming sequence in the BPT diagram.

 The [\ion{O}{3}]-detected H$\alpha$ Dots are separated into star-forming (solid green dots) and AGN (green triangles) components.  The two classes of objects are cleanly separated in Figure \ref{fig:diag}.  In all cases included in the plot we made use of the G3 spectra that provide measurements of the [\ion{N}{2}]$\lambda$6583/H$\alpha$ ratio.  The AGN are all classified as Seyfert 2 galaxies, and all show largish [\ion{O}{3}]$\lambda$5007/H$\beta$ ratios ($\geq$6.0).  Due to the selection effects associated with our survey, we would not expect to detect systems with lower equivalent-width emission lines, such as LINERs \citep{heckman1980}.   The [\ion{O}{3}]-detected star-forming galaxies all possess high-excitation spectra (with one exception).  As mentioned above, their spectra are similar to the Green Pea galaxies \citep[e.g.,][]{gp, brunker2020} that were originally recognized by their green colors and compact morphologies during visual inspection of the SDSS survey images \citep{galzoo}.  Their locations in the BPT diagram imply that many of them possess low metal abundances, which is typical of the Green Pea galaxies.

One aspect of the distribution of objects in Figure \ref{fig:diag} that is worth pointing out is the large spread in the ``excitation" values ([\ion{O}{3}]$\lambda$5007/H$\beta$ ratio) of the star-forming objects.  That is, for a given value of the [\ion{N}{2}]$\lambda$6583/H$\alpha$ ratio, the spread in [\ion{O}{3}]$\lambda$5007/H$\beta$ is larger than is typically seen in samples of ELGs.  For example, comparison with Figure 1 in either \citet{hirschauer2018} or \citet{wegner2019}, based on ELGs from the KISS objective-prism survey \citep{salzer2000,salzer2001,gronwall2004}, shows that many of the H$\alpha$ Dots extend to much lower values of [\ion{O}{3}]$\lambda$5007/H$\beta$.   This is typically interpretted as an age effect, where lower excitation values indicate ionization from older stellar populations where the most massive O stars have already ended their lives.  Apparently the narrow-band survey method is more sensitive than the objective-prism search method, allowing for the detection of star-forming regions with a wider range of ages and properties.

\subsection{Classification of Object Types}

Based on the combination of spectroscopic and morphological information, we classify each of the H$\alpha$ Dots into one of the following classes of objects.  Our classification scheme follows closely that used in \citet{hadot1}.  Here we better define what we mean by each ELG class.  The adopted classifications for each H$\alpha$ Dot are given in the second column of Table~\ref{tab:spec}.

\noindent{\bf Dwarf Star-Forming Galaxies}:  The majority of the H$\alpha$-detected sources fall into the broad category of star-forming dwarf galaxies.  As mentioned above, our selection criteria tend to restrict the choice of candidate H$\alpha$ emitting objects to be low-luminosity systems.  The compactness criterion further tends to select low redshift H$\alpha$ Dots that could be classified as Blue Compact Dwarfs (BCDs).  

Visual inspection of the sample reveals that most, but not all, of the H$\alpha$-detected Dots have properties consistent with the BCDs.  There are a few notable exceptions.  In a handful of cases (e.g., H$\alpha$ Dots 60, 127, 147) our narrow-band survey detected an \ion{H}{2} region in a low surface brightness dwarf galaxy that was too faint to be visible in our short exposure R-band continuum images.   While these are {\it bona fide} star-forming dwarf galaxies they are not compact.  Two other outliers in our sample are H$\alpha$ Dots 41 and 67.  While these two galaxies exhibit very compact morphologies with extremely strong central emission, deeper broad-band imaging (e.g., SDSS) reveals that they have a more substantial underlying host galaxy than do most of the other H$\alpha$ Dots.  Both have M$_R$ $\sim$ $-$18.5, making them by far the most luminous of the H$\alpha$-detected Dots.  For comparison, these absolute magnitudes lie approximately midway between the R-band luminosities of the Large and Small Magellanic Clouds.  

With these exceptions aside, the vast majority of the H$\alpha$-detected sources are BCD-like in nature, and may well constitute one of the best samples of such objects with distances out to $\sim$100 Mpc currently available.  The locations of many of these galaxies in the BPT diagram (Figure~\ref{fig:diag}) suggests that at least some of these BCDs are well past the peak in their star-forming episode.

\noindent{\bf Outlying/Isolated \ion{H}{2} Regions}: The H$\alpha$-detected objects that do not fall into the star-forming dwarf galaxies category are all classified as outlying or isolated \ion{H}{2} regions \citep[e.g.,][]{rw2004, werk2010}.  These are star-forming regions located beyond what appears to be the outer edge of their host galaxy.  

We classify objects as outlying \ion{H}{2} regions when they appear to be located close to a large galaxy and have brightnesses comparable to the brightest \ion{H}{2} regions visible in that galaxy.  They typically are very compact emission regions.  While we will preliminarily assign an object to this category based on the survey imaging data, spectroscopic confirmation that the \ion{H}{2} regions have velocities similar to their hosts is required before the classification can be formally adopted.  These objects usually distinguish themselves as having much higher metal abundances than are typically seen in dwarf galaxies with comparable luminosities (see \S 5.4).  

In a few cases, multiple outlying \ion{H}{2} regions have been detected in the same galaxy.  H$\alpha$ Dots 130 and 132-134 are all associated with the face-on barred spiral NGC 765, and are located well outside the main disk of the galaxy.  H$\alpha$ Dots 103 -- 106 are located in the faint outer ring of UGC 4599, which is the nearest Hoag-type ring galaxy known \citep{finkelman2011}.

\noindent{\bf Green Pea-like Galaxies}: As mentioned earlier, the H$\alpha$ Dots survey is quite sensitive to galaxies with strong [\ion{O}{3}]$\lambda$5007 emission in the redshift range 0.31 -- 0.35.  The [\ion{O}{3}]-detected galaxies are typically Seyfert 2 galaxies (N=16) or star-forming galaxies with spectra similar to the so-called Green Pea (GP) galaxies (N=40) like those in \citet{gp}.

In order to distinguish between these two classes, we typically require the G3 spectra that provide access to the redshifted H$\alpha$ and [\ion{N}{2}] lines.  G3 spectra exist for all but one of the [\ion{O}{3}]-detected galaxies discussed in the current study (the one exception is H$\alpha$ Dot 72), so that our current classifications are largely unambiguous.  We note that none of the [\ion{O}{3}]-detected SFGs shown in Figure~\ref{fig:diag} have log([\ion{O}{3}]/H$\beta$) values above 0.9, while none of the Seyfert 2 galaxies have log([\ion{O}{3}]/H$\beta$) less than 0.75.  Hence, one can typically get a good inkling of the correct class of each [\ion{O}{3}]-detected H$\alpha$ Dot without access to the [\ion{N}{2}]$\lambda$6583/H$\alpha$ ratio, but the latter line ratio is essential for definitive classifications.

Many of the [\ion{O}{3}]-detected SFGs have properties consistent with current samples of GP galaxies \citep[e.g.,][]{gp,brunker2020} and are {\it bona fide} GPs.  Others are less extreme in terms of their properties (e.g., lower star-formation rates, weaker spectra) and may simply be strongly star-forming galaxies at these intermediate redshifts.  This will become more evident in \S 5.4 below.  Hence, we will refer to these [\ion{O}{3}]-detected SFGs as GP-like galaxies.  Additional analysis of these objects is currently underway and will be reported in a separate study.

\begin{deluxetable*}{ccccccc}
\tabletypesize{\footnotesize}
\tablewidth{0pt}
\tablecaption{H$\alpha$ Dots: Average Properties within ELG Classes \label{tab:classes}}

\tablehead{
 \colhead{ELG Class} &  \colhead{Number} & \colhead{$\langle$z$\rangle$} & \colhead{$\langle$m$_R$$\rangle$} & \colhead{$\langle$M$_R$$\rangle$} & \colhead{$\langle$SFR$\rangle$} & \colhead{$\langle$log(O/H)+12$\rangle$} \\
 (1)  & (2)  & (3)  & (4)  & (5)  & (6)  & (7)
}
 \startdata
    Dwarf SFGs & 73 &  0.0154 &  18.73 &   $-$15.46 &   0.073 &  8.07  \\
    Outlying \ion{H}{2} Regions & 12 &  0.0128 &  21.00 &   $-$12.64 &   0.029 &  8.50  \\
    Green Pea-like Galaxies & 40 &  0.3416 &  20.68 &   $-$20.73 &   19.19 &  8.09  \\
    Seyfert 2s & 17 &  0.3625 &  19.76 &   $-$21.79 &   -- &  --  \\
    Seyfert 1s & 3 &  0.4987 &  19.35 &   $-$22.97 &   -- &  --  \\
    QSOs & 9 &  3.1678 &  19.78 &   $-$27.42 &   -- &  --  \\
    \enddata
\end{deluxetable*}

\noindent{\bf Seyfert 2 Galaxies}: Due to the selection technique employed by this survey, all of the Seyfert 2 galaxies are [\ion{O}{3}]-detected sources at intermediate redshifts (plus one [\ion{O}{2}]-detected galaxy at z = 0.792).  Due to the compactness criterion mentioned above, no lower redshift H$\alpha$-detected sources could fall within this classification - nearby Seyfert galaxies are all large, extended systems.

The Seyfert classification method is primarily a spectroscopic one, based on emission-line ratios that indicate photo-ionization from a nonstellar source of UV photons \citep[e.g.,][see Figure~\ref{fig:diag}]{bpt,vo1987}.  As with the GP-like galaxies, the existence of the G3 spectra is vital for unambiguous classification.   While we possess G3 spectra for all of the Seyfert 2 galaxies, we note that the single [\ion{O}{2}]-detected Seyfert 2, H$\alpha$ Dot 161, has its H$\alpha$ and [\ion{N}{2}] lines redshifted beyond the range of the G3 spectral coverage.   In this case its classification rests primarily on the observed [\ion{O}{3}]/H$\beta$ ratio.  The log([\ion{O}{3}]/H$\beta$) value for this source is 0.82, which places it in the ambiguous region where the Seyfert 2 and high-excitation GP-like galaxies overlap.  Given that the GP-like galaxies with such high [\ion{O}{3}]/H$\beta$ ratios tend to have weak [\ion{O}{2}] lines, we have tentatively classified this particular galaxy as a Seyfert 2 given that its [\ion{O}{2}] lines were strong enough to allow detection.

\noindent{\bf Seyfert 1 Galaxies and QSOs}: A primary classification criterion of Seyfert 1 galaxies and QSOs is the existence of broad emission lines of permitted transitions such as the Balmer series lines and Helium lines.  In addition, QSOs are usually distinguished by their high optical luminosities.  Because the forbidden metal lines and permitted lines emanate from different regions of the AGN they are typically not plotted in BPT-like diagrams (hence their absence from Figure~\ref{fig:diag}).

Once again, the compactness criterion used in selecting H$\alpha$ Dots eliminates the possibility of detecting any Seyfert 1 galaxies in the H$\alpha$-detected subsample.  Only three Seyfert 1s have been identified by the survey to date: one is [\ion{O}{3}]-detected (z = 0.3234), one is [\ion{O}{2}]-detected (z = 0.7941), and the third is H$\beta$-detected (z = 0.3785).   The latter source is the only H$\alpha$ Dot found to date that is detected via the H$\beta$ line.  It is included in the [\ion{O}{3}]-detected group in the redshift histogram shown in Figure~\ref{fig:redshifts}.  Naturally, all of the Seyfert 1 galaxies exhibit broad permitted lines.  Measured velocity widths (FWHM) for the three H$\alpha$ Dot Seyfert 1s range between 2300 and 5000 km/s.

The nine QSOs detected so far in the survey have all been detected via broad UV emission lines as detailed in \S 5.1.  Their reshifts range from z = 1.35 (Mg II detection) to z = 4.49 (Ly$\alpha$ detection).  Example spectra a displayed in Figure~\ref{fig:specqso}.

We summarize the object classifications of the complete set of H$\alpha$ Dots in Table~\ref{tab:classes}. The table includes the number of objects discovered within each grouping (column 2) plus average values within each group for various observed and derived quantities: redshift (column 3), apparent R magnitude (column 4), absolute R magnitude (column 5), star-formation rate (column 6) and metal abundance (column 7).   The latter two quantities are naturally only presented for star-forming systems.  We refer to the contents of this table in the following subsection.

\subsection{Properties of the H$\alpha$ Dots}

In this subsection we illustrate the distribution of derived properties for the H$\alpha$ Dots and relate these results with the previous discussions of spectral properties and ELG classifications.

\begin{figure}[h]
\epsscale{1.2}
\plotone{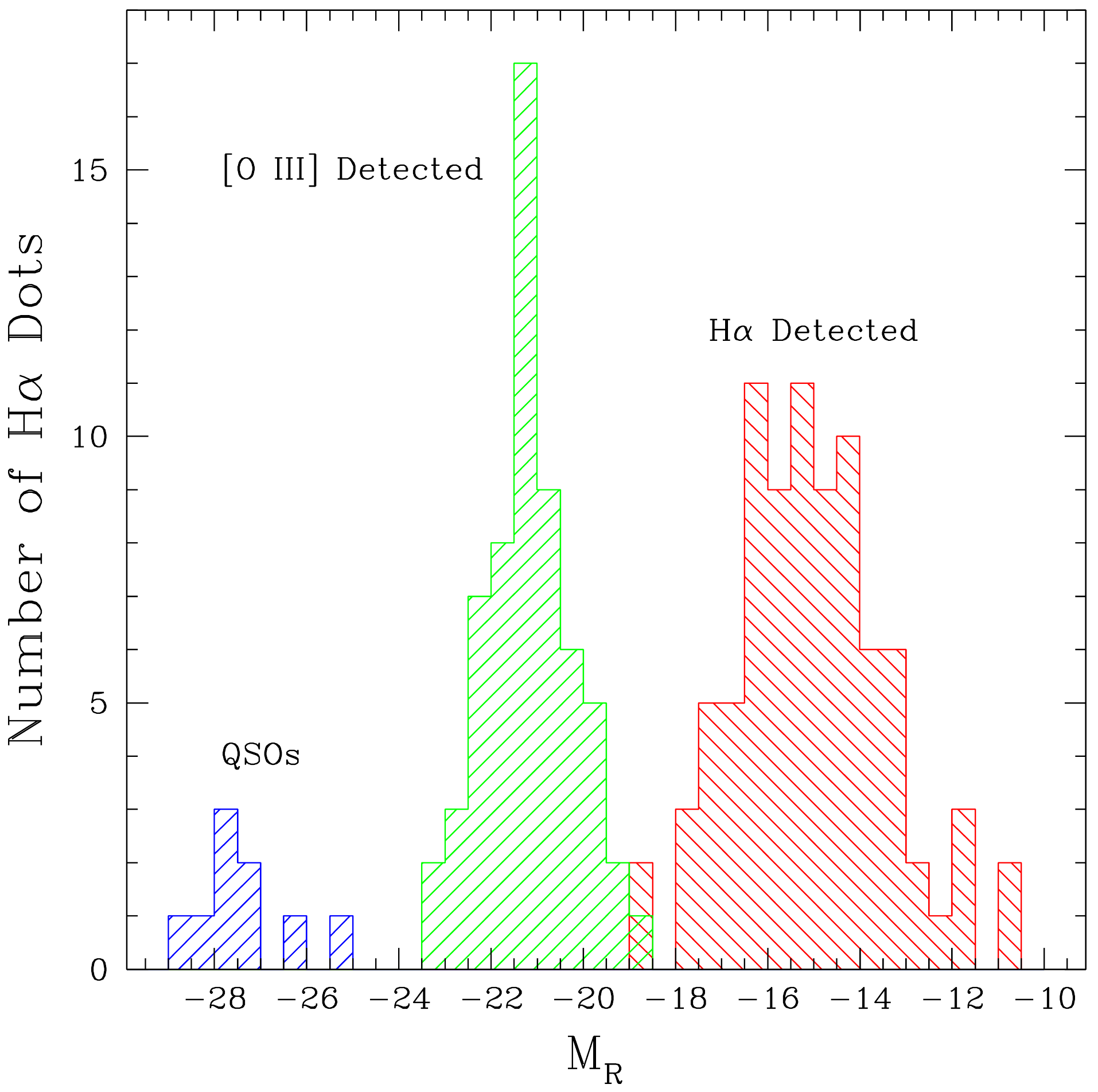}
\caption{\small Distribution of R-band absolute magnitudes for the first two lists of H$\alpha$ Dots (N=154).  Separate histograms are presented to the H$\alpha$-detected Dots (red, N=85), which are a mix of dwarf compact star-forming galaxies and out-lying \ion{H}{2} regions, the [\ion{O}{3}]- and [\ion{O}{2}]-detected objects (green, N=60), which include both Seyfert and Green Pea-like galaxies, and UV-line-detected QSOs (blue, N=9).   The latter group all have absolute magnitudes more luminous than $-$25.}
\label{fig:absmag}
\end{figure}

Figure~\ref{fig:absmag} presents the distribution of R-band absolute magnitudes for the full sample of 154 unique H$\alpha$ Dots (i.e., duplicates and false detections excluded).  The histograms are broken into three groups: H$\alpha$-detected in red, which includes the dwarf star-forming galaxies and outlying \ion{H}{2} regions; all intermediate redshift objects detected via [\ion{O}{3]} or [\ion{O}{2}] lines (plus the single H$\beta$ detection) shown in green, which includes all of the Seyfert galaxies as well as the Green Pea-like objects; high redshift objects detected by one of the strong UV emission lines in blue, which encompasses the QSOs found in the survey.

Even a cursory glance at Figure~\ref{fig:absmag} reveals that the selection method used for the H$\alpha$ Dots survey leads to the creation of well-defined, essentially non-overlapping subsamples of the strong-lined galaxy population.  The H$\alpha$-detected objects are all lower luminosity systems, which is dictated by the low redshifts covered by the survey filters for the H$\alpha$ line combined with the compactness requirement.  As seen in Table~\ref{tab:classes}, the luminosities of the outlying \ion{H}{2} regions trend toward the lowest luminosities in the sample, while the star-forming dwarf galaxies, which make up the largest subset of the H$\alpha$ Dots, exhibit a mean absolute magnitude nearly 3 magnitudes more luminous.  

The intermediate-redshift galaxies (green histogram) is almost completely separate from the H$\alpha$-detected systems in Figure~\ref{fig:absmag}, reflecting the high redshifts of the objects detected.  The Seyfert galaxies dominate the more luminous end of the distribution here, with mean M$_R$ of $-$22.97 for the three Seyfert 1s and $-$21.79 for the 17 Seyfert 2s.   The Green Pea-like galaxies cover a broad luminosity range, but dominate the numbers in the green histogram at values below $-$21.  The simple fact that such a large fraction of the H$\alpha$ Dots are [\ion{O}{3}]-detected at these intermediate redshifts speaks both to the sensitivity of the survey method (the average m$_R$ for the Green Pea-like galaxies is 20.7) and to the importance of this class of galaxy at these redshifts (see also \citealp{brunker2020}).

As one would expect, the QSOs stand apart from the other objects in the catalog in terms of their luminosities.  Since our sample of QSOs is a small one, it is not possible to draw strong conclusions regarding it.  Our sense is that, while the derived absolute magnitudes for the QSOs are consistent with other such objects, the QSOs found among the H$\alpha$ Dots tend to be among the more luminous.  This would presumably be a result of the fact that, at a given redshift, only the more luminous examples would be detectable with our survey method given the small size of our telescope.

Next we consider the H$\alpha$ line luminosities and star-formation rates (SFRs) for the H$\alpha$ Dots.  In all cases, the line luminosities are derived from the calibrated narrow-band imaging fluxes rather than from spectroscopic fluxes.  This should result in more accurate SFR estimates, since no corrections for aperture effects are required.  For the [\ion{O}{3}]-detected sources, the total H$\alpha$ flux is derived using the observed narrow-band emission-line flux combined with the spectroscopic H$\alpha$/[\ion{O}{3}] ratio.  The line fluxes for the H$\alpha$-detected galaxies are corrected for the presence of [\ion {N}{2}] emission within the filter bandpass, using the observed spectroscopic [\ion {N}{2}]/H$\alpha$ ratio; in most cases this correction is small due to the weakness of the Nitrogen lines in the low-metallicity dwarfs.   No similar correction is required for the [\ion{O}{3}]-detected objects.  The H$\alpha$ line fluxes are also corrected for absorption using the Balmer reddening coefficients (c$_{H\beta}$) listed in Table~\ref{tab:spec}.  SFRs are computed from the derived H$\alpha$ luminosities using the standard \citet{kennicutt1998} relationship:
\begin{equation}
    SFR\ [M_\odot\ yr^{-1}] = 7.9 \times 10^{-42} L_{H\alpha}.
\end{equation}
\noindent The values for L$_{H\alpha}$ and SFR for the star-forming H$\alpha$ Dots are listed in the last two columns of Table~\ref{tab:prop}.

\begin{figure}[t]
\epsscale{1.2}
\plotone{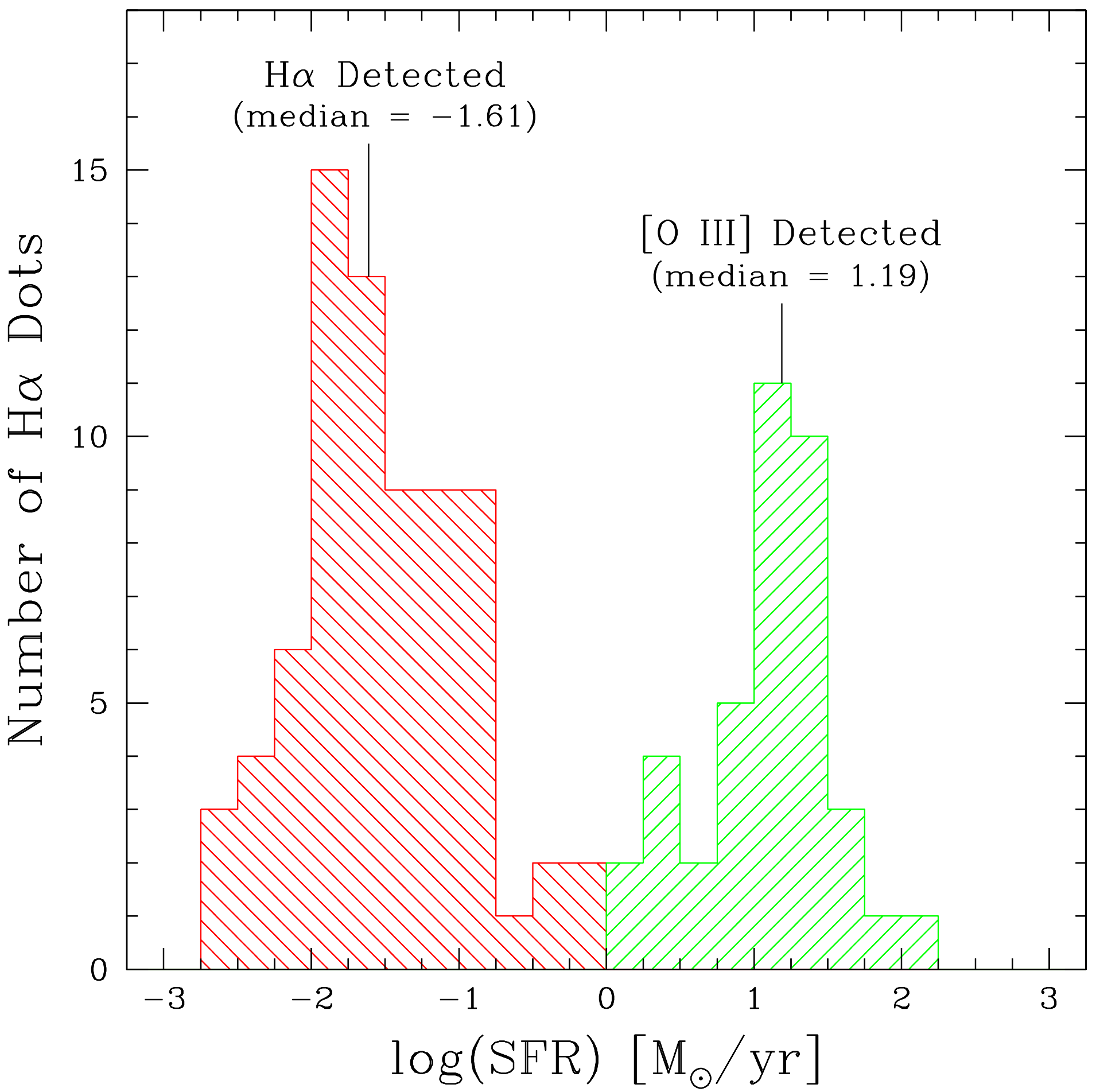}
\caption{\footnotesize Distribution of star-formation rates (SFRs) for the star-forming component of the H$\alpha$ Dots.  The red histogram shows the SFR distribution for the low-redshift H$\alpha$-detected galaxies (N=73), which are predominantly dwarfs. The SFRs of the [\ion{O}{3}]-detected Dots are shown in the green histogram (N = 39), which includes systems with SFRs in excess of 100 M$_\odot$/yr.  The median values of the two samples are indicated; they differ by a factor of 630. The [\ion{O}{3}]-detected Dots include many Green Pea galaxy candidates.}
\label{fig:sfr}
\end{figure}

The H$\alpha$-based SFRs for all of the star-forming galaxies are shown in Figure~\ref{fig:sfr}.  Once again, the lower-redshift H$\alpha$-detected galaxies are shown as the red histogram, while the intermediate-redshift [\ion{O}{3}]-detected Dots are shown in green.  For this figure, we do not include the SFRs of the outlying \ion{H}{2} regions, but rather focus on the {\it bona fide} galaxies.  As was seen in the histogram of the absolute magnitudes, the SFR distributions of the H$\alpha$- and [\ion{O}{3}]-detected galaxies segregate completely in Figure~\ref{fig:sfr}.  

The median SFR of the dwarf star-forming galaxies is 0.024 M$_\odot$\ yr$^{-1}$, substantially less the average value of 0.073 M$_\odot$\ yr$^{-1}$ reported in Table~\ref{tab:classes}.  Whichever number is chosen to parameterize the characteristic SFR of the dwarf star-forming galaxies, it is clear that the typical values are on the high side for galaxies with these types of luminosities.   This should be not surprise, given the way that the sample is selected.  Comparison with the volume-limited 11HUGS survey \citep[][see their Figure 6]{lee2009} shows that a large majority of the latter sample of galaxies with M$_B$ $>$ $-$16 (which corresponds roughly to M$_R$ $>$ $-$17) have SFRs less than 0.01 M$_\odot$\ yr$^{-1}$.  Fully 19\% of the H$\alpha$ Dots star-forming dwarfs in Figure~\ref{fig:sfr} have SFRs above 0.1 M$_\odot$\ yr$^{-1}$, compared to 0\% of the galaxies with M$_B$ $>$ $-$16 from the 11HUGS sample.  Hence, as we suggested earlier, the dwarf SFG population cataloged by our survey method is effective at discovering many strongly star-forming dwarfs located within the boundaries of the AHA survey fields, even though these objects were not themselves AHA targets.  Many of the dwarf SFGs could be classified as BCDs.

The [\ion{O}{3}]-detected SFGs exhibit substantially higher SFRs compared with the dwarfs.  The SFRs range from 1-100+ M$_\odot$\ yr$^{-1}$, with a median value of 15.5 M$_\odot$\ yr$^{-1}$.   The median SFR for the Green Pea-like galaxies is a factor of 630 times higher than the median for the dwarf SFGs!  The range of SFR values found for previously published samples of Green Pea galaxies falls between log(SFR) of 0.5 and 1.8 \citep{gp, brunker2020}. Hence, the majority of the [\ion{O}{3}]-detected H$\alpha$ Dots that are SFGs are likely to be {\it bona fide} Green Peas.

The abundances for the H$\alpha$ Dots are derived utilizing the so-called O3N2 strong-lined abundance method \citep[e.g.,][]{alloin1979,pettinipagel2004,perezmonterocontini2009,marino2013}.  Here we adopt the O3N2 calibration derived by \citet{hirschauer2018} for the KISS ELGs.  Since the latter sample consists almost exclusively of high-excitation SFGs, the \citet{hirschauer2018} calibration is only valid for galaxies close to the locus of high-excitation points that lie on the star-forming sequence in the BPT diagram (see Figure 1 of \citealp{hirschauer2018}).   This locus of points is well approximated by the theoretical curve from \citet{DE86} included in Figure~\ref{fig:diag}.  This curve in fact represents the highest excitation models computed by these authors.  

Following the prescription presented in \citet{hirschauer2018}, we identify the H$\alpha$ Dots that are located sufficiently far from the high-excitation locus that their metallicity estimates would be questionable.   These are the low-excitation objects that are located below the blue-dashed line in Figure~\ref{fig:diag}.  Most of the star-forming objects that exhibit low-excitation spectra are dwarf star-forming galaxies, although one [\ion{O}{3}]-detected source and a few outlying \ion{H}{2} region are also in this category.  We do not report a metallicity estimate in Table~\ref{tab:spec} for these objects.
In the end we derived reliable metallicity estimates for 35 dwarf SFGs, 10 outlying \ion{H}{2} region, and 36 Green Pea-like galaxies.

\begin{figure}[t]
\epsscale{1.2}
\plotone{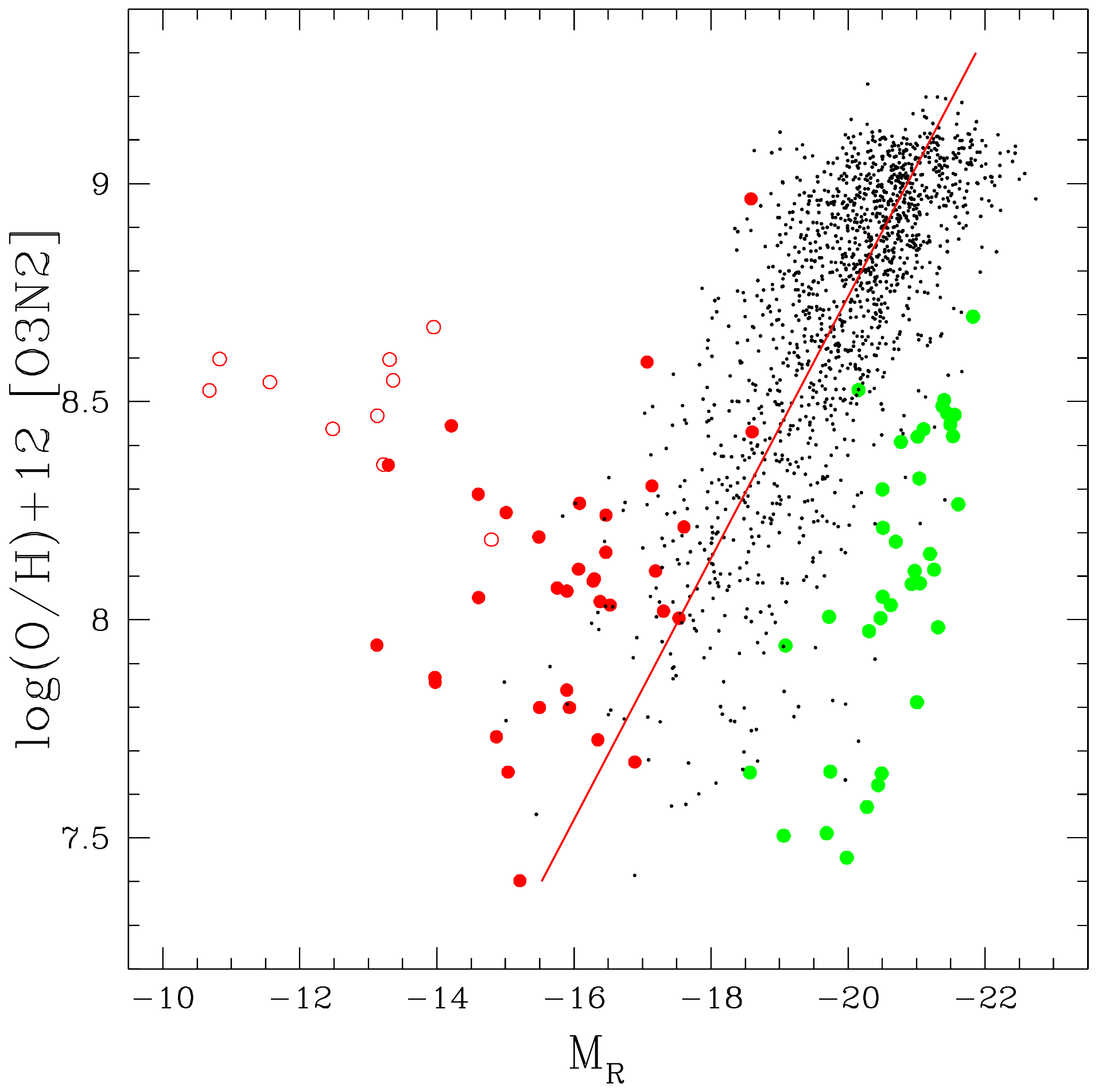}
\caption{\footnotesize Luminosity-Metallicity (L-Z) plot for the H$\alpha$ Dots.  The colored symbols have the same meaning as those used in Figure~\ref{fig:diag}.  The small block dots are star-forming galaxies from KISS (N=1450), while the solid red line is a linear fit to the KISS L-Z relation.  Both the H$\alpha$ Dots designated as outlying \ion{H}{2} regions and the [\ion{O}{3}]-detected Dots are substantially offset from the KISS L-Z relation, as discussed in the text.}
\label{fig:lz}
\end{figure}

A luminosity-metallicity (LZ) relation plot for the star-forming H$\alpha$ Dots is shown in Figure~\ref{fig:lz}.   The symbols have the same meaning as in Figure~\ref{fig:diag}: filled red circles are H$\alpha$-detected dwarf SFGs, open red cirlces are outlying \ion{H}{2} regions, and filled green circles are [\ion{O}{3}]-detected Green Pea-like galaxies.  In addition, we plot star-forming galaxies with measured abundances from KISS \citep{hirschauer2018}, shown as small dots.   The latter sample is utilized to represent a typical L-Z relation for actively SFGs.

Examination of Figure~\ref{fig:lz} reveals that none of the H$\alpha$ Dots galaxy subsamples closely follows the LZ relation defined by the KISS ELGs.  For at least two of the subsamples, this was entirely expected. We consider each of the subsamples in turn.

The expectation is that the measured metal abundances for the outlying \ion{H}{2} regions should reflect the metallicity level in the outer parts of their host galaxies (see \S 5.3).  Since these \ion{H}{2} regions are all associated with galaxies that are hundreds of times more luminous than the \ion{H}{2} regions themselves, their metal abundances should be quite high.  Hence, their locations in the LZ diagram should be substantially above the nominal relationship for real galaxies of similar luminosity.  This is exactly what we observe in Figure~\ref{fig:lz}.  Most of the outlying \ion{H}{2} regions shown in the LZ plot have O3N2 abundances of 8.5 $\pm$ 0.2, which place them between 1.5 and 2.0 dex above the extrapolation of the linear LZ relation fit to the KISS ELGs.

Typical Green Pea and Green Pea-like galaxies have metal abundances that are well below the LZ relation defined by normal galaxies \citep{gp, izotov2011, brunker2020}.  This is seen in Figure~\ref{fig:lz}, where the [\ion{O}{3}]-detected star-forming H$\alpha$ Dots all fall below the linear LZ fit to the KISS ELGs.   In many cases the offset to lower abundances is modest; there is plenty of overlap between the parameter space occupied by the star-forming KISS galaxies and the Green Pea-like H$\alpha$ Dots.  However, there are also a number of [\ion{O}{3}]-detected systems that exhibit extreme offsets in Figure~\ref{fig:lz}.   These are likely genuine Green Peas.  

While we expected the outlying \ion{H}{2} regions and Green Pea-like galaxies to not follow the LZ relation for the KISS ELGs, we had no reason to expect that the dwarf SFGs detected as H$\alpha$ Dots would also show an offset from the relation.  However, it is clear from Figure~\ref{fig:lz} that the dwarf SFGs are offset to higher abundances, on average, relative to the KISS LZ relation.  Since the same method is used to derive the metal abundance for all galaxies shown in Figure~\ref{fig:lz}, the offset cannot be attributed to differences in the abundance scales.  Furthermore, the median redshift of the 34 dwarf SFGs in Figure~\ref{fig:lz} is 4700 km/s.  Even with a large peculiar velocity of 300 km/s, a galaxy at the median redshift would only has its distance affected at the 6\% level, not enough to account for the significant offsets observed.  There are 27 H$\alpha$ Dots in the plot with M$_R$ $>$ $-$17, and all but one is above the linear fit line.  
The slope of the LZ fit to the KISS ELGs is dominated by the more numerous intermediate and high luminosity galaxies.  Hence, the dwarf SFGs in the H$\alpha$ Dot samples may be indicating that the slope of the LZ relation becomes shallower at lower luminosities.  This possibility was previously suggested by \citet{hadot1} and \citet{hirschauer2018}, among others.  We return to this issue in the next section.

\subsection{Example Applications for the H$\alpha$ Dots}

An extensive list of potentially interesting applications and follow-up projects for the H$\alpha$ Dots was given in \citet{hadot1}.   Here we summarize three specific projects that utilize three different subsamples of the H$\alpha$ Dots, in the hopes of illustrating the utility and versatility of the objects found in our survey.

{\bf Direct Abundances of Dwarf SFGs:}\ \ As we have highlighted throughout this document, the H$\alpha$ Dot selection method results in the detection of many low-luminosity star-forming galaxies in the volume of space covered by our narrow-band filters.  Our survey preferentially detects strong-lined galaxies, making them excellent candidates for detailed abundance work.  Our ``quick-look" spectra have neither the spectral coverage nor the depth to allow us to derive accurate metallicities with these initial confirming spectra.  However, we have been acquiring abundance-quality spectra for the most interesting targets as opportunity has allowed.  To date, we have collected spectra for roughly two dozen dwarf SFGs from the H$\alpha$ Dots lists that exhibit the temperature-sensitive [\ion{O}{3}]$\lambda$4363 line.   Analysis of these spectra is well under way, and will result in the determination of ``direct" abundances for a large sample of dwarf H$\alpha$ Dots (A. Hirschauer et al., in preparation).   Armed with these more accurate abundances, we plan to explore more vigorously the question of the possible slope change in the LZ relation referred to above.   

{\bf Properties of Green Pea Galaxies:}\ \ It is abundantly clear that the H$\alpha$ Dots survey is an excellent source of large numbers of previously unrecognized Green Pea galaxies \citep{gp}.  As recent studies have shown, these objects are important for our understanding of extreme star-forming systems of the type that may be responsible for contributing to the re-ionization of the universe at z $>$ 6 \citep[e.g.,][]{henry2015,izotov2016a,izotov2016b,izotov2017,izotov2018,izotov2020,jaskot2017,jaskot2019,verhamme2017,yang2017}.  In addition, the Green Peas appear to be among the most rapidly evolving galaxy population observed \citep{brunker2020}.  Hence studying their properties at as many redshifts as possible is important.

The H$\alpha$ Dots survey detects GPs in the redshift range 0.31 -- 0.35, which represents the outer limit of the \citet{gp} sample.  Members of the H$\alpha$ Dots survey team have begun detailed studies of these interesting and enigmatic objects.  We are obtaining high quality spectra for direct abundances for many of the GP candidates.   In addition, we are exploring the spatial distributions of galaxies in the vicinity of these GPs with the goal of understanding their environments (S. Brunker, in preparation).  We note that the H$\alpha$ Dots survey represents an emission-line flux-limited sample of galaxies with a quantifiable completeness limits.   Hence, we will be able to derive reliable volume densities of the GP galaxies, allowing us to better understand their evolution with lookback time as well as their contribution to the total SFR density at their observed redshifts.

{\bf Volume Densties and Abundanaces of Intermediate Redshift Seyfert 2 Galaxies:}\ \ The [\ion{O}{3}]-detected H$\alpha$ Dots also includes a well-defined sample of Seyfert 2 galaxies located in the same redshift range as the GPs discussed above.  Once again, the nature of the survey will allow us to measure accurate volume densities of strong-lined Seyfert 2s in the z = 0.31 -- 0.35 range.   This redshift range lies beyond the limits of current wide-field redshift surveys (e.g., none of the 17 H$\alpha$ Dot Seyfert 2s has been observed spectroscopically by SDSS), allowing us to push the census of strong-lined AGN to larger distances.  In addition, preliminary evidence suggests that at least some of the [\ion{O}{3}]-detected Seyfert 2 galaxies {\bf may} have significantly lower metal abundances than their low redshift counterparts.   This is seen in Figure~\ref{fig:diag}, where a number of the H$\alpha$ Dot Seyfert 2s have lower values of [\ion{N}{2}]/H$\alpha$ than are typically observed in local systems.  For example, none of the large sample of nearby Seyfert 2s studied by \citet{carvalho2020} have log([\ion{N}{2}]/H$\alpha$) below $-$0.4, while 6 of 16 of the H$\alpha$ Dot Seyfert 2s have values below this limit.  We are currently exploring methods for deriving reliable abundances for Seyfert 2 galaxies (D. Carr, in preparation), with an eye toward evaluating any possible evolution with lookback time.

\section{Summary \& Conclusions}

We present a second list of point-like emission-line objects known as H$\alpha$ Dots.  These objects were discovered serendipitously by carrying out systematic searches of hundreds of narrow-band images obtained for the ALFALFA H$\alpha$ project \citep{vansistine2016}.

A total of 354 fields were searched, resulting in the discovery of 119 compact/unresolved emission-line candidates (112 unique objects after accounting for duplications due to field overlap).  Our imaging data yield R-band magnitudes and calibrated emission-line fluxes for all ELG candidates.  In addition, all of our candidates have been observed spectroscopically.  We present redshifts and emission-line diagnostics for all of our newly-detected objects.  These spectra reveal the nature of the objects being cataloged.   In particular, they show that our survey method is sensitive to the detection of objects via a variety of emission lines: H$\alpha$, [\ion{O}{3}]$\lambda$5007, H$\beta$, [\ion{O}{2}]$\lambda$3727, and several UV emission lines commonly observed in QSOs, including Ly$\alpha$.  

We combine and analyze all 180 H$\alpha$ Dots from the first two survey lists to develop a more comprehensive picture of  the make-up of our ELG sample.  We are able to categorize this apparently heterogeneous group of strong-lined sources into just five classes of galaxies with activity: dwarf star-forming galaxies (including many blue compact dwarfs), outlying/isolated \ion{H}{2} regions, Grean Pea-like galaxies, Seyfert 2 galaxies, and broad-line AGN (mostly QSOs).  Our narrow-band selection method concentrates these classes into specific redshift windows, depending on the emission line that is used in their selection.   This means that the individual H$\alpha$ Dot categories represent samples of both line-flux-limited and redshift-limited galaxies.  We utilize the available observational data to explore the properties of the galaxies in each of these groups

The H$\alpha$ Dots survey project is ongoing and continues to be carried out primarily by undergraduate research students.  Additional lists of H$\alpha$ Dots are being prepared for publication (e.g., D. Watkins in preparation), and our program of follow-up spectroscopy is continuing.  We are also pursuing detailed studies of specific subsamples of the H$\alpha$ Dots, as highlighted in \S 5.5.   In addition to generating more survey lists of new ELG candidates, future papers in this series will derive detailed metallicities of the dwarf star-forming H$\alpha$ Dots, explore the properties of the Green Pea galaxies found in the survey, and analyze the population of intermediate redshift Seyfert 2 galaxies uncovered by this work.

\acknowledgements

We thank the anonymous referee for making numerous suggestions that greatly improved the presentation of this paper.
We gratefully acknowledge the financial support of the College of Arts and Sciences and the Department of Astronomy at Indiana University, which made our access to the WIYN 0.9-m telescope possible.  The H$\alpha$ Dots survey project is based on data obtained for the ALFALFA H$\alpha$ project, which was carried out with the support of the National Science Foundation (NSF-AST-0823801).  

This project made use of Sloan Digital Sky Survey data.  Funding for the SDSS and SDSS-II has been provided by the Alfred P. Sloan Foundation, the Participating Institutions, the National Science Foundation, the U.S. Department of Energy, the National Aeronautics and Space Administration, the Japanese Monbukagakusho, the Max Planck Society, and the Higher Education Funding Council for England. The SDSS Web Site is http://www.sdss.org/.  The SDSS is managed by the Astrophysical Research Consortium for the Participating Institutions. The Participating Institutions are the American Museum of Natural History, Astrophysical Institute Potsdam, University of Basel, University of Cambridge, Case Western Reserve University, University of Chicago, Drexel University, Fermilab, the Institute for Advanced Study, the Japan Participation Group, Johns Hopkins University, the Joint Institute for Nuclear Astrophysics, the Kavli Institute for Particle Astrophysics and Cosmology, the Korean Scientist Group, the Chinese Academy of Sciences (LAMOST), Los Alamos National Laboratory, the Max-Planck-Institute for Astronomy (MPIA), the Max-Planck-Institute for Astrophysics (MPA), New Mexico State University, Ohio State University, University of Pittsburgh, University of Portsmouth, Princeton University, the United States Naval Observatory, and the University of Washington.
\facilities{WIYN:0.9m, HET(LRS)}

\end{document}